\pdfoutput=1
\documentclass[structabstract]{aa}  
\usepackage{graphicx}
\usepackage{txfonts}
\usepackage{natbib}
\usepackage{bm}
\usepackage[colorlinks=true,linkcolor=blue,citecolor=blue,filecolor=blue,urlcolor=blue]{hyperref}
\bibpunct{(}{)}{;}{a}{}{,} 
\begin{document}

\title{Application of the coherent structure tracking to solar Doppler maps 
  to determine horizontal velocity fields at the Sun's surface}

\author{ M. Sampoorna\inst{1}\fnmsep\thanks{\email{sampoorna@iiap.res.in}} 
\and T. Roudier\inst{2} \and F. Paletou\inst{3}}

          \institute{Indian Institute of Astrophysics, Koramangala, Bengaluru
  		560034, India
            \and
          Cnrs,  Institut de Recherche en Astrophysique et
            Plan\'etologie, 14 av. E. Belin, F--31400 Toulouse, France
            \and
	    Universit\'e de Toulouse, Observatoire Midi-Pyr\'en\'ees,
          Cnrs, Cnes, Irap, Toulouse, France
            }

   \date{Received May 30, 2025; accepted May 22, 2026.}

   \abstract
    {Coherent Structure Tracking (CST) is a technique for determining 
    the solar surface horizontal flows at high spatial and 
    temporal resolution by tracking the proper motion of granules. CST has 
    been traditionally applied to solar intensity images in the continuum, 
    which clearly depict the granular patterns. However, solar granulation is 
    also visible in the Dopplergrams.} 
   {We aim to show that CST can be applied to solar Dopplergrams to derive 
   the solar surface horizontal velocity fields with the same level of 
   confidence as those determined by CST on intensity images.}
   {For this purpose, we apply the CST to continuum intensity 
   images and Dopplergrams obtained from SDO/HMI and also from a numerical 
   simulation of granulation. We then compare the 
   resulting solar surface horizontal velocity 
   fields and their derivatives (namely, the horizontal divergence 
   and the vertical component of the vorticity) for different 
   time windows.} 
   {Pearson's linear global correlation coefficient (GCC) 
   between the horizontal velocity fields determined from CST on 
   Doppler and on intensity images of a relatively less active Sun is 
   about 73\% for a 30 minutes time average, while the 
   corresponding local correlation coefficient (LCC) near the disk center is 
   about 80\%. For the divergence of the horizontal 
   velocity field, we obtain a GCC of 72\% and a near disk center LCC 
   of 84\%. The curl of the horizontal velocity field being more 
   noisy exhibits somewhat reduced GCC and LCC. 
   These coefficients increase with increasing time 
   window. A similar trend is exhibited by 
   Spearman's and Kendall's rank-order correlation coefficients, 
   although they are somewhat smaller in value. 
   The different correlation coefficients slightly decrease 
   for magnetically more active Sun with sunspots or 
   emerging pores in a plage 
   region. A high correlation is obtained between the 
   horizontal flows derived by applying CST to intensity and vertical 
   velocity maps from a numerical simulation.}
   {}


   \titlerunning{Coherent Structure Tracking applied to solar Doppler maps}

   \maketitle
   \nolinenumbers

\section{Introduction}
\label{sec-intro}
It is well-known that most of the solar phenomena occurring in its atmosphere 
are the result of interplay between the solar surface flows and the magnetic 
fields anchored in the convection zone. Hence, it is crucial to understand and 
determine the nature of solar surface flows from small to large scales. 
While Doppler provides a direct determination of line-of-sight 
component of plasma velocity on the Sun, indirect methods are needed to 
determine the transverse component. Many different techniques exist to 
measure this transverse or horizontal component of the solar surface flows, 
such as local correlation tracking \citep[LCT;][]{nov86}, feature tracking 
\citep{sl94,sl95}, time-distance helioseismology \citep{djh93}, Fourier 
local correlation tracking \citep[FLCT;][]{fw08}, and DeepVel 
\citep{rametal17}, to name a few. Coherent structure tracking 
\citep[CST;][]{tretal12} is a method that 
tracks the proper motions of solar granules to determine the surface 
horizontal flows. 
Basically CST uses solar granules as passive scalars to follow the underlying 
plasma motions \citep[see][for more details on CST technique]{tretal99,cstnew07,tretal12,rinetal17,tretal18}.
This technique can be used to study short to long term flows (such as solar 
differential rotation and meridional circulation) on the surface of the Sun \citep{mahaetal24}.

It is important to note that the different flow methods mentioned 
above (excepting the time-distance helioseismology which uses consecutive 
Dopplergrams) determine the solar surface horizontal flows using consecutive 
continuum images of the Sun. Thus they give an estimation of the optical 
flows at the solar surface and may not represent the actual solar plasma 
motions. Thus several studies in the literature have focused on comparing 
the horizontal velocities derived from these flow methods with the 
actual plasma velocities obtained from the state-of-the-art magnetohydrodynamic 
simulations \citep[see e.g.,][]{rieuetal01,vermaetal13,degetal14,louisetal15,rametal17,tremetal18}. 
\citet{rieuetal01} show that granules are able to trace statistically the 
large-scale flows (at meso- and super-granular scales) but lead to a 
systematic underestimation of the actual velocities. Indeed they 
show that horizontal velocities determined from CST miss the actual 
plasma velocities roughly by a factor of two 
\citep[see Fig.~4 of][]{rieuetal01}. \citet{vermaetal13} 
show that LCT determined horizontal velocities are underestimated roughly 
by a factor of three. We remark that while CST follows individual granules, 
LCT uses spatial window that includes both granules and intergranular lanes. 
Thus the LCT measurements are affected by the width of the spatial window 
\citep[see e.g.,][]{vermaetal13,louisetal15}. \citet{louisetal15} show 
that the horizontal velocities retrieved from LCT lack the fine structure 
observed in the simulated data, although there is an overall morphological 
agreement between the two. \citet{tremetal18} show that DeepVel 
reproduces the horizontal velocities at subgranular and granular 
scales very well and is second only to FLCT at mesogranular and 
supergranular scales, although it is dependent on the simulation model used 
to train the neural network. Time-distance helioseismology is also shown 
to underestimate the solar surface horizontal flows \citep{degetal14}. 
Despite this shortcoming of underestimation of flow velocities by different 
flow techniques there are no other method available for an accurate 
and exact determination of actual solar plasma motions at all scales.
Therefore, underestimation of flow velocities continue 
to remain also for the studies presented in this paper using CST on 
Doppler maps. 

As already mentioned above, the CST technique is traditionally 
applied to time sequences of the solar intensity images in the continuum, 
such as those provided for every 45 sec by the Helioseismic and Magnetic Imager 
\citep[HMI:][]{hmi112,hmi212} 
onboard the Solar Dynamics Observatory \citep[SDO:][]{sdo12}. 
In this technique the velocities are 
measured by following the trajectory of each granule (a coherent structure) 
over the course of its life time (which is defined as the time between its 
birth and death). Measured horizontal velocities and the Doppler 
velocities from HMI Dopplergrams are then corrected for satellite  
motion, limb effect (namely limbshift correction applied to Doppler velocities 
to account for geometry induced errors), and the tilt of the solar rotation 
axis with respect to the observer \citep[see Appendix A of][]{rinetal17}. 
In this way CST provides the solar surface velocity field 
($V_x$, $V_y$) in a Cartesian coordinate system, where $x$ and 
$y$ are in the plane of the sky with $x$ increasing toward solar west (or 
parallel to the direction of the solar rotation) and $y$ increasing toward 
solar north, and $z$ is directed toward the observer along the line-of-sight. 
The horizontal velocity fields ($V_x$, $V_y$) can then be combined 
with the line-of-sight velocity field $V_z$ determined from HMI Dopplergrams to 
construct the photospheric full vector velocity field in spherical coordinates 
\citep{tretal13,rinetal17}. CST provides the horizontal flow fields with a 
spatial resolution of 2.5 Mm and a temporal resolution of 30 min. 

In this paper, we show that the CST technique can be applied to 
Dopplergrams\footnote{In this context, we remark that 
\citet{tretal19} have shown that the LCT can be applied to Doppler maps 
obtained from simulation or observations to determine horizontal velocity 
fields which are in very good agreement with those obtained by applying the 
LCT to intensity images.} to determine $V_x$ and $V_y$, as solar 
granulation pattern is seen in Dopplergrams too. For this purpose, 
we consider Dopplergrams obtained from SDO/HMI as well as from a numerical 
simulation of solar granulation. Using global and local correlation 
coefficients, such as Pearson's linear correlation and Spearman's 
and Kendall's rank correlation statistics, we compare the surface flow 
fields obtained by applying the CST to intensity and Doppler maps. The 
comparison is also made for the horizontal divergence and vertical component 
of the curl of the flow field. Furthermore, we also present these 
comparisons by considering the pixel by pixel similarity measure given by the 
mean squared error. For our comparison we consider three time averaging 
windows of 30, 60, and 120 minutes. 

\section{Data}
\label{sec-data}

\subsection{Numerical simulation}
\label{ssec-num-simu}
Following \citet[][see also \citealt{tretal19}]{tretal20} we use the 
first two hours sequence of the full 27 hours sequence of the (synthetic white 
light) emergent intensity and vertical velocity ($V_z$) field resulting from 
the 3D magneto-convection simulation of solar granulation
\citep[][see also the review by \citealt{s12}]{sn98,setal09}. We use this 
sequence at the continuum optical depth of unity (namely, $z=0$ Mm). The 
simulation data cube has the dimension of 1008 $\times$ 1008 pixels in the 
horizontal $(x,y)$ plane with the pixel size of 79.26 km and a time 
step or time cadence of $\Delta t = 
60$ sec. Since the CST is tailor made for applying to the SDO/HMI intensity 
images which has the pixel size of $0^{\arcsec}\!\!.5$, we had to rebin the 
simulation pixel to $0^{\arcsec}\!\!.5$ to be able to apply the CST to the 
simulation data. 

\subsection{Observations}
\label{ssec-obs}
In this paper we use the data from SDO/HMI. More specifically, we 
make use of the full-disk continuum images of the solar photosphere 
near the Fe\,{\sc i} 6173\ \AA\ absorption line and the maps of the 
line-of-sight velocity (Dopplergrams) observed by SDO/HMI on 29th 
of December 2022, 28th and 29th of May 2023. The data on 29th of December 2022 
corresponds to a relatively less active Sun (with only a few spots at higher 
latitudes somewhat closer to the limb), while those on 28th and 29th of May 
2023 corresponds to magnetically more active Sun 
with two large sunspots close to the equator and a plage region showing 
emerging spots near the equator, respectively. These sets of data were chosen 
to verify the suitability of using CST on Doppler maps to determine horizontal 
flow fields when compared to CST on intensity. SDO/HMI records these data for 
every 45 sec with a pixel scale of 
$0^{\arcsec}\!\!.5$. We apply the CST algorithm on these data sets for the 
first 30, 60, and 120 minutes to derive the horizontal velocity fields and 
their derivatives (such as horizontal divergence $D=\partial_x V_x + 
\partial_y V_y$ and vertical component of the vorticity 
$\zeta = \partial_x V_y - \partial_y V_x$). In the case of data 
from 29th of May 2023, we apply the CST algorithm to first 10 hours of the 
observed data, to understand the nature of horizontal surface 
flows in the plage region with emerging spots or pores.

\section{Comparison of horizontal surface flows from CST on intensity and Doppler maps}
\label{sec-intvsdop}
In this section we compare the horizontal flow fields and their 
derivatives determined by applying the CST technique to intensity images 
and to Doppler maps. We remark that application of CST to Doppler maps 
does not require any modification to the CST algorithm. One simply provides 
Dopplergrams as input instead of the intensity images. While the application 
of CST to intensity requires processing Doppler maps to obtain 
line-of-sight velocities for constructing the full 
three-dimensional velocity vector, the 
application of CST to Doppler maps fully avoids 
processing intensity images, thereby saving 
a considerable amount of data storage and computing time 
involved in processing intensity images in the first 
step of CST\footnote{For the CST code and for more details on 
the CST method see the CST manual available at 
\url{https://idoc.osups.universite-paris-saclay.fr/medoc/tools/cst-codes/}}.
As mentioned in Section~\ref{sec-data} we apply the CST technique 
to three temporal sequences of data from a numerical simulation of 
granulation and observations from SDO/HMI.

\begin{figure*}[]
\sidecaption
\includegraphics[width=12cm]{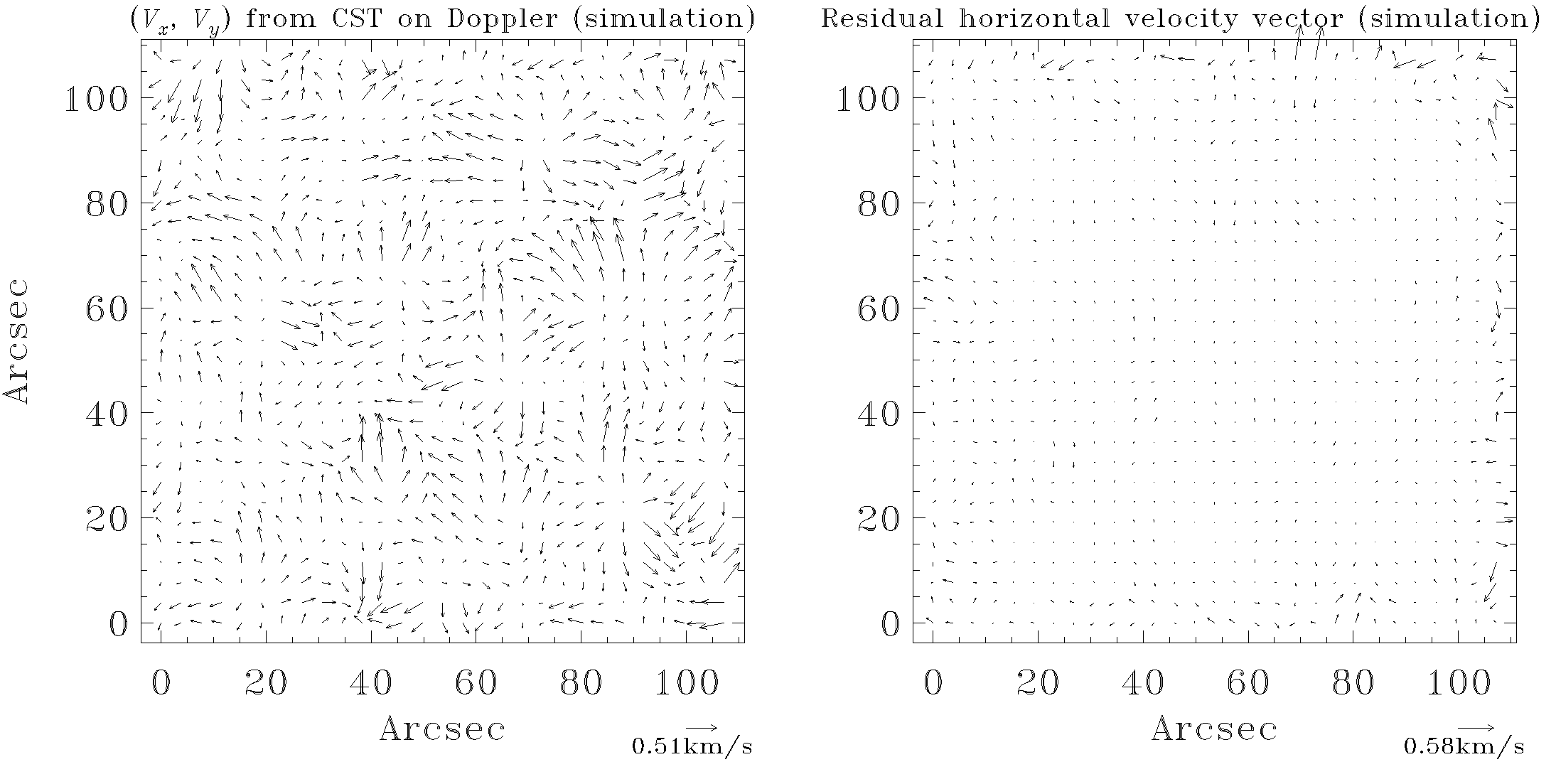}
\caption{
	Left\,: Horizontal velocity vector ($V_x$, $V_y$) determined 
	by applying the CST to vertical velocity ($V_z$) map obtained from a 
	numerical simulation of granulation. Right\,: The residual 
	vector between the horizontal velocities obtained by applying the CST 
	to intensity and vertical velocity map from simulation. A temporal 
	window of 30 min is used. The arrow to the bottom right corner of 
        each panel shows the length of the maximum horizontal velocity 
	vector in the left panel and the maximum residual vector in the right 
	panel. Although they are of similar magnitude, the residual vector is 
        considerably smaller at all the places except near the edges.
	  }
  \label{dopvsint-velovect-simu}
\end{figure*}

\begin{table*}
\caption{Pearson's linear, and Spearman's and Kendall's rank 
correlation coefficients between different flow quantities determined by 
applying the CST to intensity and vertical velocity ($V_z$) maps 
from a numerical simulation of granulation.} 
\label{simu-table}      
\centering                          
\begin{tabular}{c c c c c c c c c c}        
\hline\hline                 
Flow  & \multicolumn{3}{c}{Pearson's linear}
& \multicolumn{3}{c}{Spearman's rank-order}
	& \multicolumn{3}{c}{Kendall's rank-order}
	\\    
Quantity  & \multicolumn{3}{c}{correlation coefficient}
& \multicolumn{3}{c}{correlation coefficient}
	& \multicolumn{3}{c}{correlation coefficient}
	\\    
                    & (30 min) & (60 min) & (120 min)  
	 & (30 min) & (60 min) & (120 min)  
	 & (30 min) & (60 min) & (120 min)  
	\\    
\hline                        
$V_x$ & 0.934 & 0.962 & 0.989
	 & 0.930 & 0.957 & 0.989 
	 & 0.805 & 0.844 & 0.938 \\      
$V_y$ & 0.915 & 0.959 & 0.980 
	 & 0.917 & 0.960 & 0.984 
	 & 0.786 & 0.850 & 0.924 \\      
$V_{\rm amp}$ & 0.880 & 0.921 & 0.961
		  & 0.876 & 0.926 & 0.965 
		  & 0.721 & 0.779 & 0.887 \\      
Divergence $D$ & 0.892 & 0.947 & 0.975 
		  & 0.904 & 0.950 & 0.978 
		  & 0.765 & 0.823 & 0.903 \\      
Curl $\zeta$ & 0.820 & 0.871 & 0.949 
		& 0.821 & 0.875 & 0.955 
		& 0.663 & 0.714 & 0.856 \\      
\hline                                   
\end{tabular}
\tablefoot{Each of these coefficients are 
given for 30, 60, and 120 minutes temporal windows.}
\end{table*}

Apart from presenting a visual comparison of the horizontal flow 
fields obtained by CST on intensity and Doppler maps, we present a more 
quantitative comparison via the Pearson's linear correlation coefficient 
and the rank-order correlations such as the Spearman's and Kendall's 
rank correlation coefficients. The Pearson's linear correlation coefficient 
is defined as 
\begin{equation}
C_{ab} = {\frac{\sum_{i=1}^n (a_i-\bar a)(b_i-\bar b)}
{\sqrt{\left[\sum_{i=1}^n(a_i-\bar a)^2\right]\left[\sum_{i=1}^n(b_i-\bar b)^2\right]}}},
\label{pcc}
\end{equation}
where $n$ is the size of the sample, $a_i$, $b_i$ are the individual sample 
points, with $\bar a$ and $\bar b$ representing their mean, respectively. In 
our case $a$ and $b$ could be $V_x$, $V_y$, $V_{\rm amp}=\sqrt{V_x^2+V_y^2}$, 
horizontal divergence $D$, or curl $\zeta$ obtained by applying CST to 
intensity and Doppler maps, respectively. The Spearman's 
rank correlation coefficient between the data sets $a_i$ and $b_i$ is simply 
the Pearson's correlation between the rank values of those two data sets. 
Unlike the Pearson's correlation which gives the measure of linear 
relationships between the two data sets, Spearman's rank correlation is a 
measure of monotonic relationships between the data sets (irrespective of if 
it is linear or not). In other words it indicates the correlation between the 
two sets of ranks. Like the Spearman's rank correlation, Kendall's rank 
correlation also measures the monotonic relationships between the two data 
sets, but its calculation is based on the concordant or discordant pairs. More 
specifically, Kendall's rank correlation coefficient is the ratio of difference 
in number of concordant and discordant pairs to total number of pairs. Here 
a pair $(a_i,b_i)$ and $(a_j,b_j)$ is said to be concordant if the sort 
order of $(a_i,a_j)$ and $(b_i,b_j)$ agrees (i.e., if either both $a_i > a_j$ 
and $b_i>b_j$ holds or both $a_i < a_j$ and $b_i<b_j$ holds), otherwise the 
pair is said to be discordant. Clearly the computing time for Kendall's 
rank coefficient increases with sample size $n$. Hence, it is usually computed 
for smaller data sets. Furthermore, to measure the image similarity pixel by 
pixel, we also compute the mean squared error (MSE) defined as 
\begin{equation}
{\rm MSE}_{ab} = {\frac{1}{n}} \sum_{i=1}^n (a_i-b_i)^2.
\label{mse}
\end{equation}

\subsection{Comparison using simulation data}
\label{sec-comp-simu}
Using numerical simulation, \citet{rieuetal01} have shown that 
the horizontal velocity fields derived by tracking the proper 
motion of granules are highly correlated with the actual velocity 
fields of the plasma for meso and supergranular scales, although 
a systematic underestimation of actual flows persist. For this purpose, 
they apply both CST and LCT to the intensity images from the numerical 
simulation and compare the resulting horizontal velocities with the 
actual flow velocities. They show that results from both the techniques 
broadly agree with the numerical simulation. 
Hence, here we focus on comparing the horizontal velocities obtained by 
applying the CST to intensity and vertical velocity maps derived from a 
numerical simulation (see Sect.~\ref{ssec-num-simu}). We recall 
that the time cadence of numerical simulation is 60 sec and we consider 
first two hours of the simulated data. Left panel of 
Figure~\ref{dopvsint-velovect-simu} shows the horizontal 
velocity vector obtained by applying the CST to vertical velocity ($V_z$) 
derived from this numerical simulation. A temporal window of 30 min has been 
considered. The residual velocity vector, defined as the difference between 
the horizontal velocity vectors obtained by applying the CST to intensity 
images and to vertical velocity maps from simulation is shown in the right 
panel of Figure~\ref{dopvsint-velovect-simu}. 
The latter clearly shows that the flows obtained by both the methods are 
highly similar except at the very edge (particularly the right edge). These 
differences however reduce with increasing temporal window. This can be 
clearly seen from the Pearson's linear, and Spearman's and Kendall's rank 
correlation coefficients presented in Table~\ref{simu-table} for different flow 
quantities. We have also computed the MSE (see Eq.~(\ref{mse})) for all the 
different flow quantities and for the three time windows. The maximum MSE of 
$0.0044$ was obtained for divergence $D$ with 30 min time window and the 
minimum MSE of $0.0003$ was obtained for $V_x$ with 120 min time window. 
Clearly MSE is quite small and it indicates a higher similarity.

\subsection{Comparison using observational data}
\label{sec-comp-obs}

\begin{figure*}[]
\sidecaption
\includegraphics[width=12cm]{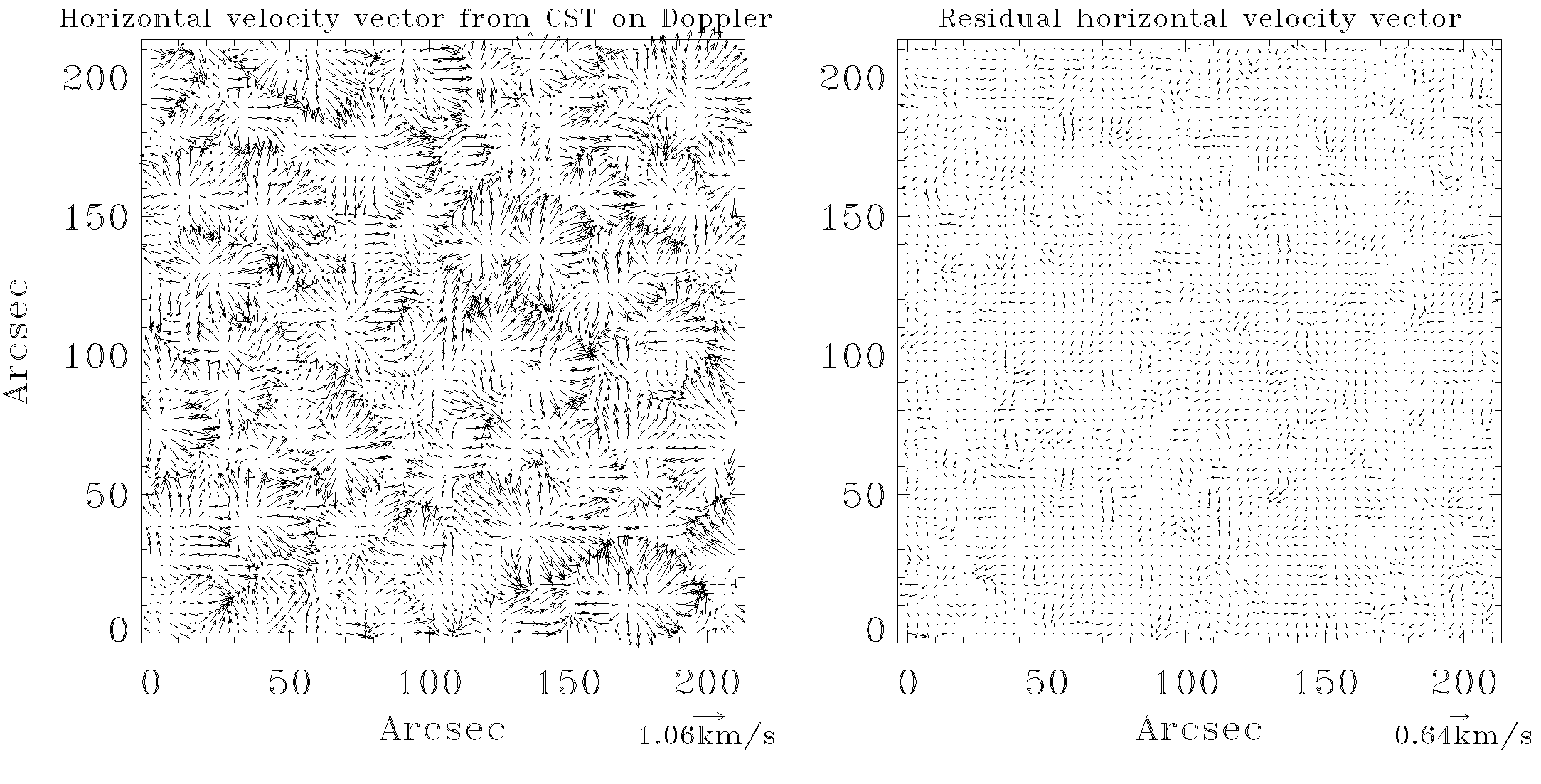}
	\caption{
        Left\,: Horizontal velocity vector ($V_x$, $V_y$) determined 
	by applying the CST to Dopplergrams from 29th of December 2022 
	observations of SDO/HMI is shown for a region near the disk center and 
	for a temporal window of 30 min. Right\,: The residual 
        vector between the horizontal velocities obtained by applying the CST
        to intensity and Doppler maps from the above-said  
	observations is shown for the same region and time window as in the 
	left panel. The arrow at the bottom right corner of both the 
        panels shows the length of the maximum horizontal velocity
        vector in the left panel and the maximum residual vector in the right
	panel.
	  }
  \label{dopvsint-velovect}
\end{figure*}

We first apply the CST to the relatively less active Sun data from 
29th of December 2022. 
The horizontal velocity vector obtained by applying the CST to Dopplergrams 
is shown in the left panel of Figure~\ref{dopvsint-velovect}. It covers 
a region of $210 \times 210$ arcseconds near the disk center. A temporal 
window of 30 min has been considered. Similarity between the flows obtained 
by applying the CST to intensity images and Dopplergrams can clearly be 
inferred from the residual velocity vector shown in the right panel of 
Figure~\ref{dopvsint-velovect}. A quantitative comparison based on the 
different correlation coefficients is presented in Tables~\ref{gcc-table} and 
\ref{lcc-table}. 

Table~\ref{gcc-table} displays the Global Correlation Coefficient 
(GCC) obtained for different flow quantities using the 30, 60, and 120 
minutes time averaging. For computing GCC, the size of the sample $n$ 
(see e.g., Eq.~(\ref{pcc})) is the full-disk of the 
Sun observed by the SDO/HMI, namely $4096 \times 4096$ pixels with 
a pixel scale of $0^{\arcsec}\!\!.5$. However, because the CST provides the 
horizontal velocity fields on a spatial scale of 2.5 Mm (about 7 
SDO/HMI pixels), $n$ for GCC would be $586 \times 586$ pixels 
(wherein now one pixel is $3^{\arcsec}\!\!.5$). The GCC 
increases with increasing time window for both Pearson's 
linear and Spearman's rank-order correlations (see Table~\ref{gcc-table}). 
For all the flow quantities excepting the curl the GCC is reasonably high 
indicating a very good correlation between the CST on intensity images and 
Doppler maps. Clearly, for the flow velocities $V_x$, $V_y$, and $V_{\rm amp}$ 
we obtain an average Pearson's linear GCC of about 73\% for 30 min 
time window, which increases 
to 81\% and 86\% for 60 and 120 minutes time windows, respectively. 
The horizontal divergence $D$ of the flow field also exhibits good 
correlation. However, the curl $\zeta$ being more noisy exhibits a reduced 
correlation which however increases with increasing time window. 
Spearman's rank-order GCC for different flow quantities exhibits a 
similar trend as the corresponding Pearson's linear GCC, although its values 
are slightly smaller than those of Pearson's linear GCC.

\begin{table*}
\caption{Pearson's linear and Spearman's rank-order GCC for 
different flow quantities with the 30, 60, and 120 minutes time averaging. }
\label{gcc-table}      
\centering                          
\begin{tabular}{c c c c c c c}        
\hline\hline                 
Flow  & \multicolumn{3}{c}{Pearson's linear GCC}  
	& \multicolumn{3}{c}{Spearman's rank-order GCC} \\    
Quantity & (30 min) & (60 min) & (120 min)
	 & (30 min) & (60 min) & (120 min)  \\    
\hline                        
   $V_x$ & 0.717 & 0.807 & 0.861 
	 & 0.665 & 0.757 & 0.819 \\      
   $V_y$ & 0.704 & 0.791 & 0.845 
	 & 0.647 & 0.733 & 0.795 \\      
   $V_{\rm amp} $ & 0.762 & 0.822 & 0.860 
		  & 0.840 & 0.875 & 0.899 \\      
   Divergence $D$ & 0.720 & 0.824 & 0.893 
		  & 0.638 & 0.742 & 0.820 \\      
   Curl $\zeta$ & 0.464 & 0.586 & 0.695 
		& 0.420 & 0.526 & 0.626 \\
\hline                                   
\end{tabular}
\tablefoot{Relatively less active Sun observations from SDO/HMI on 29th of 
December 2022 have been considered.}
\end{table*}

\begin{table*}
\caption{Pearson's linear and Spearman's and Kendall's rank-order 
LCC near disk center for different flow quantities with the 30, 60, and 
120 minutes time averaging.}
\label{lcc-table}      
\centering                          
\begin{tabular}{c c c c c c c c c c}        
\hline\hline                 
Flow  & \multicolumn{3}{c}{Pearson's linear LCC}  
      & \multicolumn{3}{c}{Spearman's rank-order LCC}  
      & \multicolumn{3}{c}{Kendall's rank-order LCC}  \\    
Quantity & (30 min) & (60 min) & (120 min)  
         & (30 min) & (60 min) & (120 min)  
         & (30 min) & (60 min) & (120 min)  \\    
\hline                        
   $V_x$ & 0.858 & 0.919 & 0.952 
	 & 0.858 & 0.918 & 0.952 
	 & 0.666 & 0.749 & 0.807 \\      
   $V_y$ & 0.859 & 0.920 & 0.952 
	 & 0.853 & 0.915 & 0.948 
	 & 0.664 & 0.747 & 0.804 \\      
   $V_{\rm amp} $ & 0.674 & 0.785 & 0.851 
	 & 0.650 & 0.769 & 0.839 
	 & 0.461 & 0.570 & 0.643 \\      
   Divergence $D$ & 0.839 & 0.913 & 0.947 
	 & 0.828 & 0.905 & 0.945 
	 & 0.635 & 0.733 & 0.796 \\      
   Curl $\zeta$ & 0.585 & 0.707 & 0.789 
	 & 0.561 & 0.684 & 0.768 
	 & 0.392 & 0.496 & 0.574 \\      
\hline                                   
\end{tabular}
\tablefoot{LCC is computed using a $210\times 210$ 
arcseconds window near the disk center. Relatively less active Sun 
observations from SDO/HMI on 29th of December 2022 have been considered.} 
\end{table*}

We also consider the local correlation coefficient (LCC) and 
study its variation across the full-disk of the relatively less active 
Sun. Figure~\ref{dopvsint-lcc} displays the 
Pearson's linear LCC for the different flow quantities obtained using 
30 (panels (a)), 60 (panels (b)), and 120 (panels (c)) minutes time window. 
For computing the LCC 
we divide the $586 \times 586$ pixels into spatial windows of $20 \times 20$ 
pixels so that $n$ (see Eq.~(\ref{pcc})) for LCC is given by the latter. We 
recall that one pixel after CST application is $3^{\arcsec}\!\!.5$. 
As with Pearson's linear GCC (see Table~\ref{gcc-table}), the LCC 
also increases with increasing time window and the curl of the horizontal 
flows exhibit a lower LCC than the other flow quantities. 

For a given time window, 
Pearson's linear LCC decreases near the limb 
which is an edge or projection effect. At high latitudes and 
longitudes (close to the limb) the granules in intensity images are grouped 
and hence when segmentation is applied close to the limb, individual granules 
are not isolated due to the curvature near the limb. As a result the proper 
motion of individual granules in intensity cannot be measured near the 
limb. This is referred to as edge or projection effect. Moreover, due 
to the center-to-limb variation of the intensity, near the limb we sample the 
granules in the higher layers than at disk center, so that the granules exhibit 
lesser contrast and appear to blur or fade near the limb. On the 
other hand in Dopplergrams, at the disk center the Doppler velocity represents 
the vertical velocity inside the granule, and close to the limb it represents 
the horizontal velocity of the inside of the granule (granule expansion). 
Hence, between the disk center and the 
limb, Doppler of granule is a mix between these two components (namely, 
vertical and horizontal velocity inside the granule). Thus when we go from 
disk center to the limb the correlation between CST on intensity and on 
Doppler is affected. Nevertheless, with the Doppler, close to the limb, we 
follow always same entities (more horizontal granule expansion) which help 
to measure the horizontal velocities $V_x$ and $V_y$ on the Sun 
\citep[and subsequently $V_r$, $V_{\phi}$, $V_{\theta}$; see e.g.,][]{tretal13}. 
Because of the above-mentioned reasons the LCC in some places 
close to the limb also becomes negative (indicating 
anti-correlation). In order to better visualize this whenever LCC is negative 
it is simply set to -1 irrespective of its actual value. As a result the 
places where LCC is negative appear as black squares in 
Fig.~\ref{dopvsint-lcc}. Thus the gray scale gradation seen in 
this figure are restricted to only positive values of LCC. 

Fig.~\ref{dopvsint-lcc} clearly demonstrates that CST applied to 
full-disk Doppler maps produces flow fields similar to those derived from 
intensity (though with slight differences). This will be useful for 
determining large-scale flows such as differential rotation and meridional 
circulation, when long time sequence of data needs to be treated (see 
Sect.~\ref{sec-conclu} for a discussion on comparison of computational 
requirement for CST on Doppler and on intensity images). Indeed CST has been 
previously applied to full-disk intensity images from SDO/HMI to derive 
large-scale flows \citep[see e.g.,][]{tretal18,mahaetal24,upetal24}. 
Other optical flow techniques such as LCT and self-supervised optical 
flow methods have also been applied to full-disk observations to 
determine the large-scale flows \citep[see][]{loptetal17,lietal23}.

In Table~\ref{lcc-table} we give for different flow quantities the 
values of Pearson's linear and Spearman's and Kendall's rank-order 
LCC computed using a $210\times 210$ arcseconds window near the disk 
center (similar to that considered for Figure~\ref{dopvsint-velovect}). 
As expected Pearson's linear and Spearman's rank-order 
LCC near disk center are larger than the corresponding GCC 
(compare Tables~\ref{gcc-table} and \ref{lcc-table}). In the case of curl 
of the flow field, both the Pearson's and Spearman's LCC near disk 
center have also improved when compared to the 
corresponding GCC. Since the computing time for the 
Kendall's rank-order correlation increases significantly with increasing 
sample size $n$, it has been computed only for the near disk center region 
(see Table~\ref{lcc-table}). Furthermore, for $V_x$, $V_y$, and $V_{\rm amp}$ 
together we obtain an average near disk center Pearson's linear LCC 
of about 80\% for 30 min time averaging, which then increases to 87\% and 
92\%, respectively for 60 and 120 minutes time averaging. 
Both the rank-order LCC are slightly smaller than the Pearson's linear 
LCC, but they increase with increasing time window like the Pearson's linear 
LCC. As for the MSE (see Eq.~(\ref{mse})), a maximum value of $0.027$ was 
obtained for $V_y$ with 30 min time window and a minimum value of $0.0058$ 
was obtained for the curl $\zeta$ with 120 min time window.

\begin{figure*}[]
\sidecaption
\includegraphics[width=12cm]{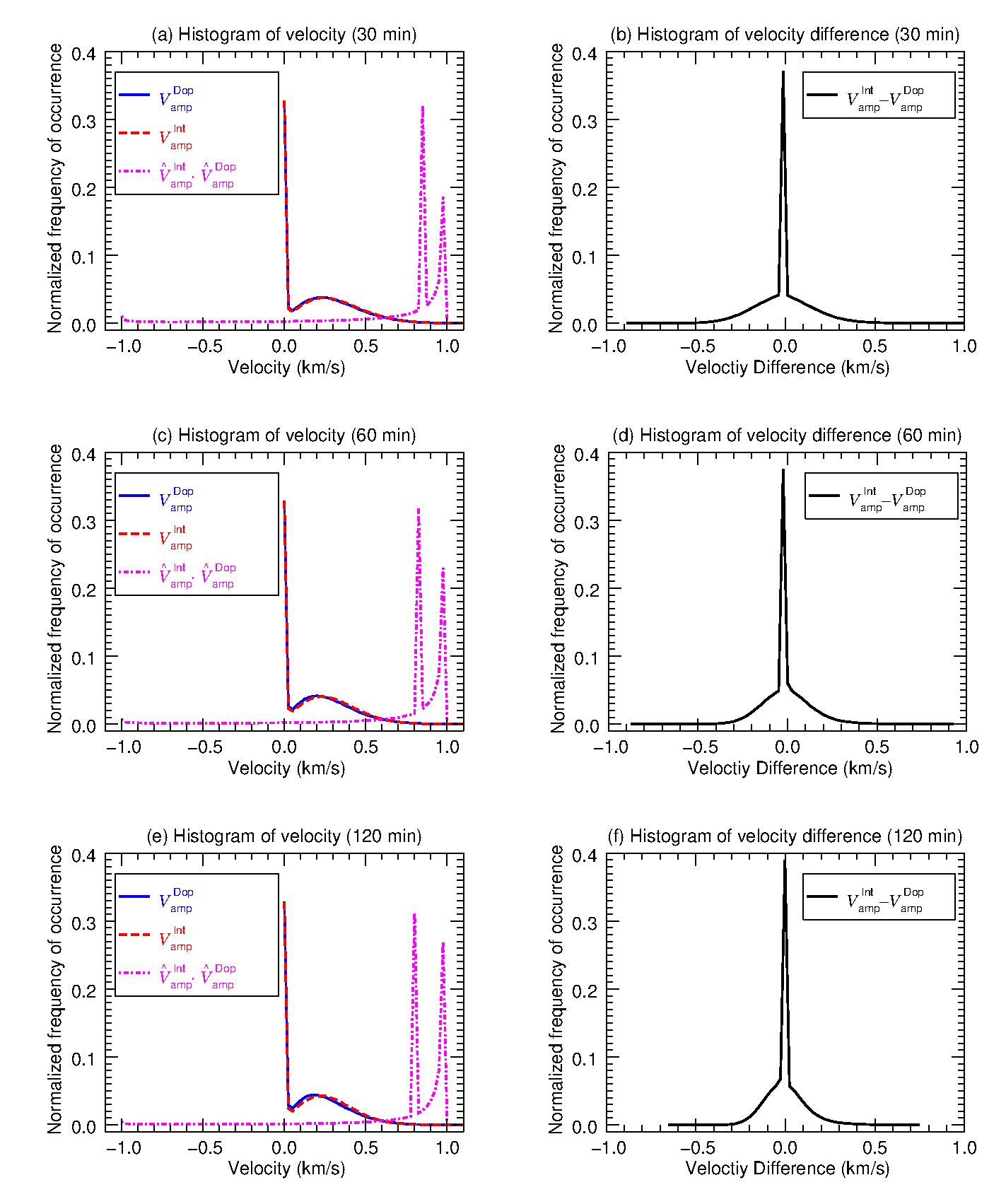}
\caption{Histogram of the amplitude of the horizontal velocity 
	field (panels (a), (c), (e)) and the difference in velocity 
	amplitude obtained from CST on intensity and 
	Dopplergrams (panels (b), (d), (f)) for 30 (panels (a) and (b)), 
	60 (panels (c) and (d)), and 120 (panels (e) and (f)) minutes time 
	window. In the left panels, we also shown the histogram of 
	the scalar product (see Eq.~(\ref{scal_prod})) between the horizontal 
	velocities obtained from CST on intensity and Doppler maps. In this 
	case the abscissa represents the cosine of the angle between 
	${\bm \hat V}^{\rm Int}$ and ${\bm \hat V}^{\rm Dop}$ which ranges 
	from $-1$ to $+1$. In both the panels, 
	the superscripts ``Int'' and ``Dop'' indicate 
	that the corresponding velocity is obtained by applying CST to 
	intensity and Dopplergrams, respectively. Note that in the left 
	panels, amplitude of the horizontal velocity $V_{\rm amp}$ 
	being a positive quantity is restricted to positive values of velocity.
	Relatively less active Sun observations from
	SDO/HMI on 29th of December 2022 have been considered.
	}
  \label{histo}
\end{figure*}

In the left panels of Figure~\ref{histo}, we present the histogram of 
the amplitude of the horizontal velocity field $V_{\rm amp}$ 
obtained by applying the CST to intensity images and Dopplergrams. We 
also present the histogram of the scalar product of horizontal velocities 
obtained from CST on intensity and Doppler maps, namely
\begin{equation}
{\bm \hat V}^{\rm Int}\cdot {\bm \hat V}^{\rm Dop}=
{\frac{V_x^{\rm Int}\,V_x^{\rm Dop}+V_y^{\rm Int}\,V_y^{\rm Dop}}{V_{\rm amp}^{\rm Int}\,V_{\rm amp}^{\rm Dop}}}, 
\label{scal_prod}
\end{equation}
where in the superscripts ``Int'' and ``Dop'' indicate that the corresponding 
velocity is obtained by applying the CST to intensity and 
Dopplergrams, respectively. 
In the right panels of Figure~\ref{histo}, we show the histogram for the 
difference in $V_{\rm amp}$ obtained by the two methods mentioned 
above. A fine resolution (or bin size) of $0.025\,{\rm km/s}$ has been 
used in the case of velocity amplitudes and $0.025$ in the case of scalar 
product. Again the histograms are shown for 30, 60, and 120 minutes time 
averaging windows. 
The high degree of similarity between the histograms obtained for 
both the methods (compare blue and red lines) is clearly visible in 
panels (a), (c), and (e), which shows that the velocity amplitudes 
are correctly determined by the CST on Dopplergrams. This is also evident 
in the velocity amplitude difference histogram, where differences 
are confined to the small velocity amplitude. This is expected 
because the identification of granules in the segmented data in intensity and 
that in Dopplergrams could be little different, which then give some 
incertitude on the location of the center of gravity of each granule. 
The granule area in CST on Doppler can also be different from granule area 
in CST on intensity. The velocity higher than 0.3 km/s seems in good 
accordance. A similar behavior as noted for $V_{\rm amp}$ is also 
seen for $V_x$ and $V_y$ individually (figure not illustrated). 
Furthermore, the histogram of the scalar 
product (see magenta line in panels (a), (c), (e)) is confined to values of 
the cosine of the angle between ${\bm \hat V}^{\rm Int}$ and 
${\bm \hat V}^{\rm Dop}$ in the range $0.8$ to $1$ indicating that the 
horizontal velocity directions are also well recovered. However, 
the histogram of the scalar product exhibits two peaks, while we expect only 
one peak around unity. The reason for the second peak around $0.8$ could be 
due to the fact that between the disk center and the limb Doppler measures 
both vertical and horizontal component of the flow velocity while CST on 
intensity measures only the horizontal component. Nevertheless, from all the 
studies presented so far 
we can conclude that the CST can be applied to the Doppler maps of 
relatively less active Sun to derive the horizontal flows.

\begin{table*}
\caption{Pearson's linear and Spearman's and Kendall's rank-order 
LCC for different flow quantities obtained with the 30, 60, and 120 minutes 
time averaging.} 
\label{lcc-sunspot-table}      
\centering                          
\begin{tabular}{c c c c c c c c c c}        
\hline\hline                 
Flow  & \multicolumn{3}{c}{Pearson's linear LCC}  
      & \multicolumn{3}{c}{Spearman's rank-order LCC}  
      & \multicolumn{3}{c}{Kendall's rank-order LCC}  \\    
Quantity & (30 min) & (60 min) & (120 min)  
         & (30 min) & (60 min) & (120 min)  
         & (30 min) & (60 min) & (120 min)  \\    
\hline                        
$V_x$ & 0.662 & 0.783 & 0.855
	 & 0.649 & 0.763 & 0.834 
	 & 0.469 & 0.574 & 0.651 \\      
$V_y$ & 0.747 & 0.839 & 0.897
	 & 0.710 & 0.804 & 0.866 
	 & 0.525 & 0.617 & 0.689 \\      
$V_{\rm amp} $ & 0.519 & 0.664 & 0.751 
	 & 0.460 & 0.607 & 0.706 
	 & 0.319 & 0.430 & 0.513 \\      
Divergence $D$ & 0.679 & 0.794 & 0.865 
	 & 0.665 & 0.780 & 0.850 
	 & 0.479 & 0.585 & 0.664 \\      
Curl $\zeta$ & 0.434 & 0.550 & 0.653
	 & 0.440 & 0.546 & 0.640 
	 & 0.302 & 0.382 & 0.459 \\      
\hline                                   
\end{tabular}
\tablefoot{A region around sunspots near to the equator with a field of 
view of $301\times 126$ arcseconds is considered. CST has been applied to 
magnetically active Sun observations from SDO/HMI on 28th of May 2023.}
\end{table*}

We next consider the application of CST to magnetically active 
Sun data from 28th of May 2023. Here we focus on a field of view of $301
\times 126$ arcseconds covering the region around two large sunspots near 
to the equator to assess the 
applicability of CST to Doppler around sunspots, although CST has been 
applied to the full Sun intensity and Doppler images. The different correlation 
coefficients between horizontal velocities derived from CST on intensity and 
Doppler maps around the sunspots are displayed in Table~\ref{lcc-sunspot-table} 
for the three different time windows. Clearly the different correlation 
coefficients are relatively smaller than those obtained for relatively 
less active Sun case (compare Tables~\ref{lcc-table} and 
\ref{lcc-sunspot-table}). This is expected as convection is suppressed in 
the sunspots and hence the horizontal velocities cannot be detected effectively 
by tracking the granules especially in the umbra. Furthermore, since the 
sunspots are not at the disk center, there exists some projection effects, 
which can modify the aspect of solar granule in Doppler and make some small 
differences on the amplitude and direction of the horizontal velocity fields. 
This is particularly true in the case of CST on Doppler, as the 
horizontal velocities here are derived by tracking the granular proper motion 
in Doppler maps, unlike in \citet{LBS13} who directly derive the horizontal 
velocities from Doppler maps of sunspots located at heliocentric angles 
of about $50^\circ$ when the line-of-sight velocities are predominantly 
horizontal. Thus while CST on Doppler (and even CST on intensity) leads to 
underestimation of horizontal velocities in the umbral and penumbral regions 
of the sunspot, the method of \citet{LBS13} does not lead to such 
underestimation especially when heliocentric angles are on the order of 
$50^\circ$. Nevertheless, all the three correlation coefficients increase 
with increasing time window, indicating that CST can be applied to Doppler 
maps around sunspots to derive meaningful horizontal flows for time windows 
of 60 min or larger. The MSE for the field of view covering the sunspots is 
also slightly larger than that obtained for the relatively less active Sun case 
considered previously. More specifically, a maximum MSE of $0.0492$ was 
obtained for $V_x$ with 30 min time window and the 
minimum MSE of $0.0111$ was obtained for curl $\zeta$ with 120 min time 
window. 

\begin{figure*}[]
\begin{center}
\includegraphics[scale=0.62]{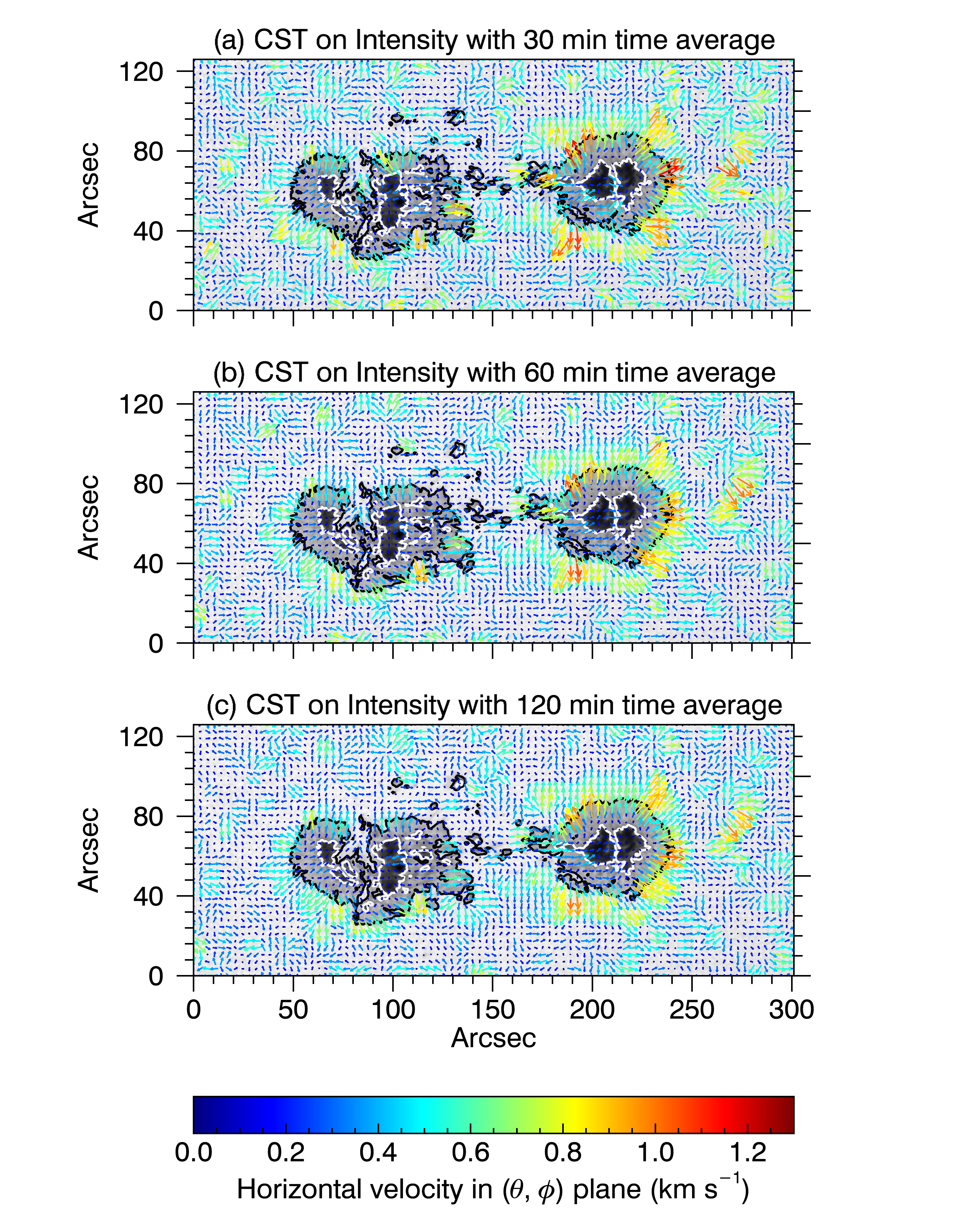}
\hspace*{-1.6cm}\includegraphics[scale=0.62]{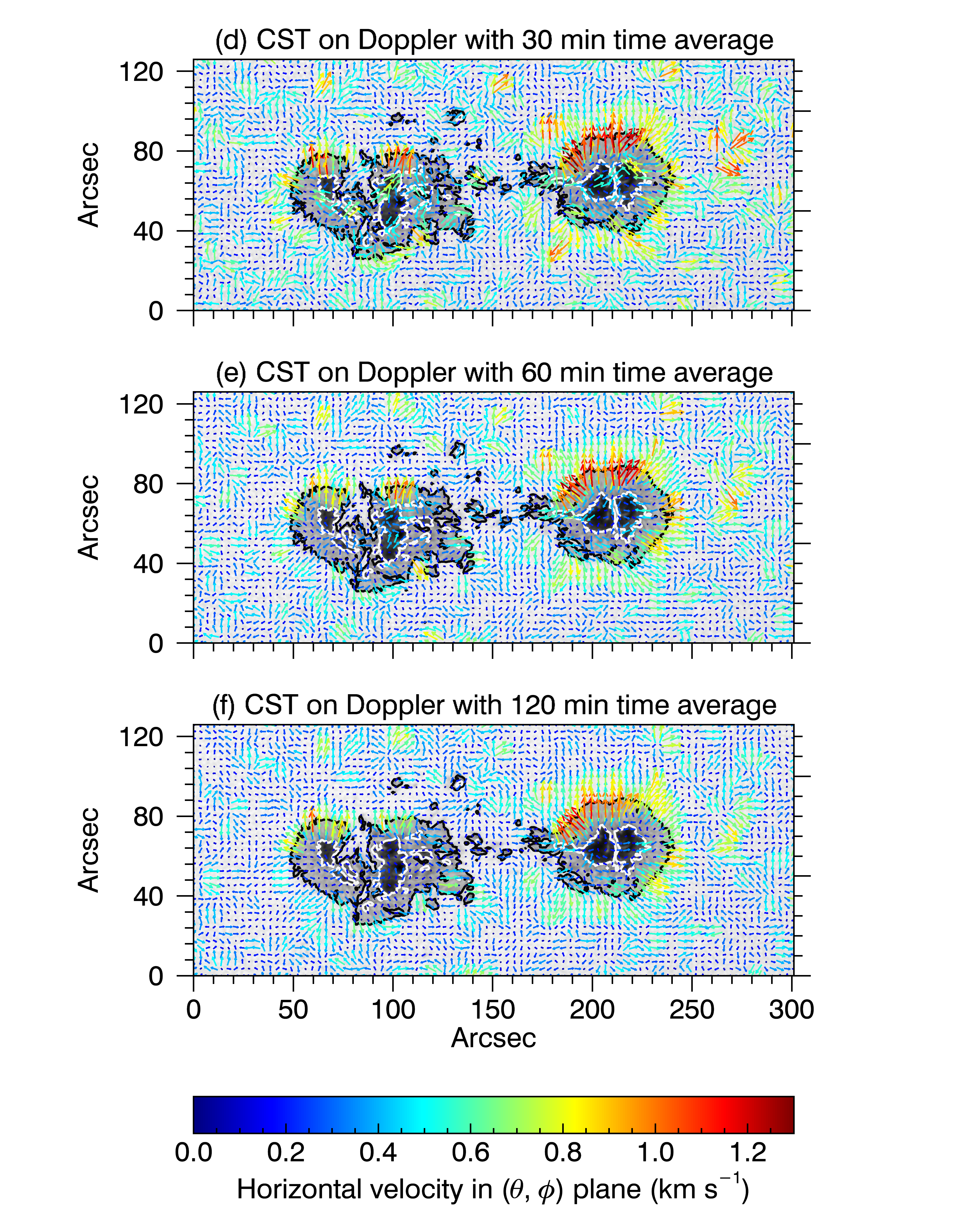}
\end{center}
	\caption{Horizontal velocity vector ($V_{\phi}$, $V_{\theta}$) 
	derived by applying CST to intensity (left column) and Doppler (right 
	column) maps observed by SDO/HMI on 28th of May 2023. Field of view 
	covers the region around the sunspots near to the equator. The top, 
	middle, and bottom rows correspond to 30, 60, and 120 min time 
	averages, respectively. Color and length of the arrows indicate 
	the horizontal velocity in a spectral scale depicted by the color bar. 
	The time averaged intensity image is displayed in the background. 
	The black and white contours mark the boundary of penumbra and  
	umbra, respectively. Notice the moat flow towards the west of 
	leading sunspot. 
	}
\label{int_dop_hvel_28may2023}
\end{figure*}

In Figure~\ref{int_dop_hvel_28may2023} we compare the horizontal 
velocities determined by applying the CST to intensity and Doppler maps around 
the sunspot region. Here we plot the horizontal velocities in the spherical 
coordinates, namely $V_{\phi}$ and $V_{\theta}$ as they represent longitudinal 
and latitudinal flows, respectively. The transformation of the 
horizontal velocities $V_x$ and $V_y$ together with Doppler velocity to 
velocities in spherical coordinates, namely, $V_r$, $V_{\phi}$, and 
$V_{\theta}$ is described in Section~5 of \citet{tretal13}. For the reasons 
already noted above, we see that the differences in horizontal velocities from 
CST on intensity and Doppler are significant mainly within the sunspots and 
to a lesser extent (only in amplitude) around the outer boundary of penumbra 
(namely the black contour), which however decreases with increasing time 
window. In general the amplitude of the horizontal velocity from CST on 
Doppler is slightly larger than that determined from CST on intensity. While 
CST on intensity persistently shows inflows in the umbra and the inner 
boundary of penumbra (see the region in and around the white contour in left 
panels of Figure~\ref{int_dop_hvel_28may2023}) for all the 
three time windows, CST on Doppler shows this only for time window of 120 min. 
Radial outflows around the outer boundary of penumbra (black contour; excepting 
the east side) in the leading sunspot (west one) is seen for all the three 
time windows and for both the methods. This basically represents the moat flow 
around the leading sunspot \citep[see e.g.,][]{vermaetal18}. More 
specifically, the horizontal velocity in the west side of the leading sunspot 
is pointed towards the west. This is expected as the leading sunspot moves 
more quickly than the solar rotation unlike the trailing sunspot. We remark 
that the moat flow is more developed in the leading sunspot than in the 
trailing sunspot. Using FLCT on a 6-year time series of 
continuum images from SDO/HMI, \citet{loptetal17} show that large-scale 
inflows around active regions converge predominantly towards the trailing 
polarity. This may perhaps explain the suppressed moat flow around 
the tailing polarity observed in Fig.~\ref{int_dop_hvel_28may2023}. 
Furthermore, Outflows are also seen starting a bit inside 
the edge of the penumbra to the periphery of penumbra in both the sunspots 
especially in the north and south side of the sunspots. This is seen in all 
the three time windows and from both the methods.

\begin{figure*}[]
\sidecaption 
\includegraphics[width=6cm]{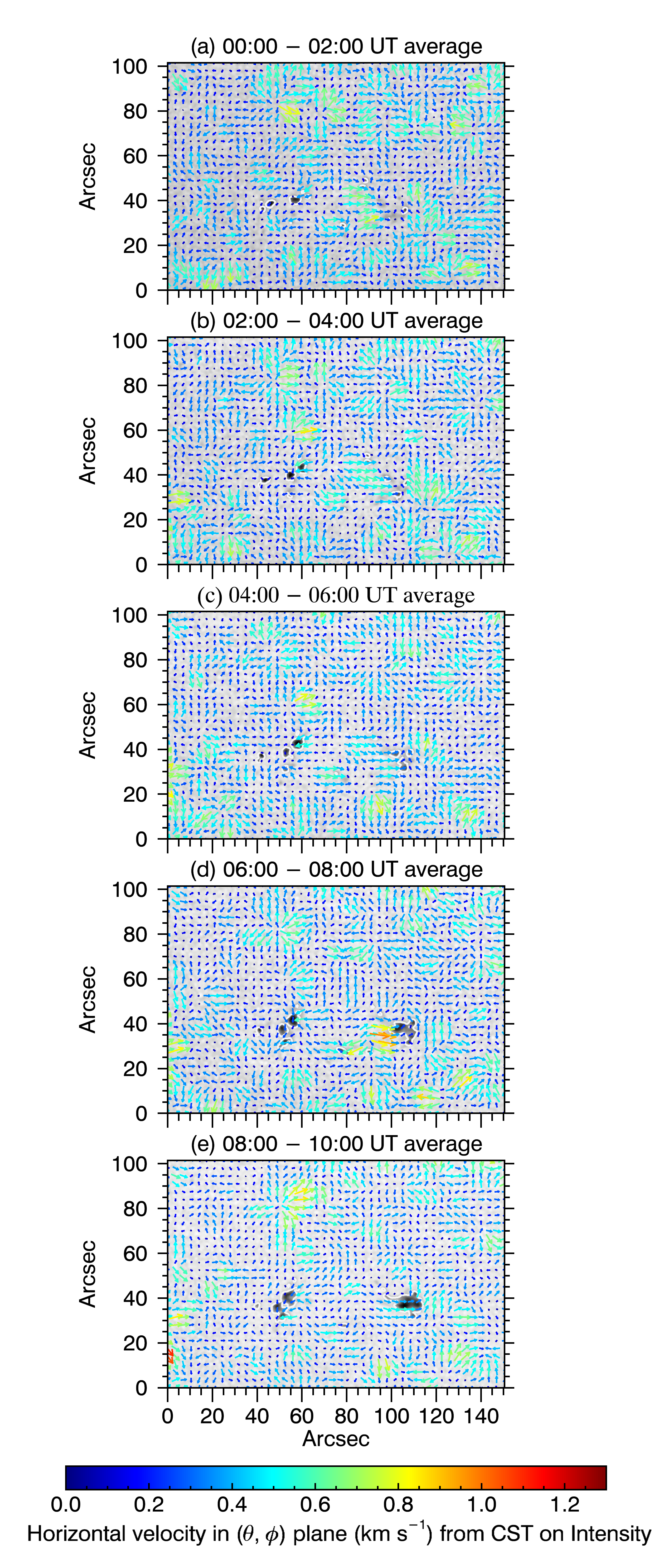}
\includegraphics[width=6cm]{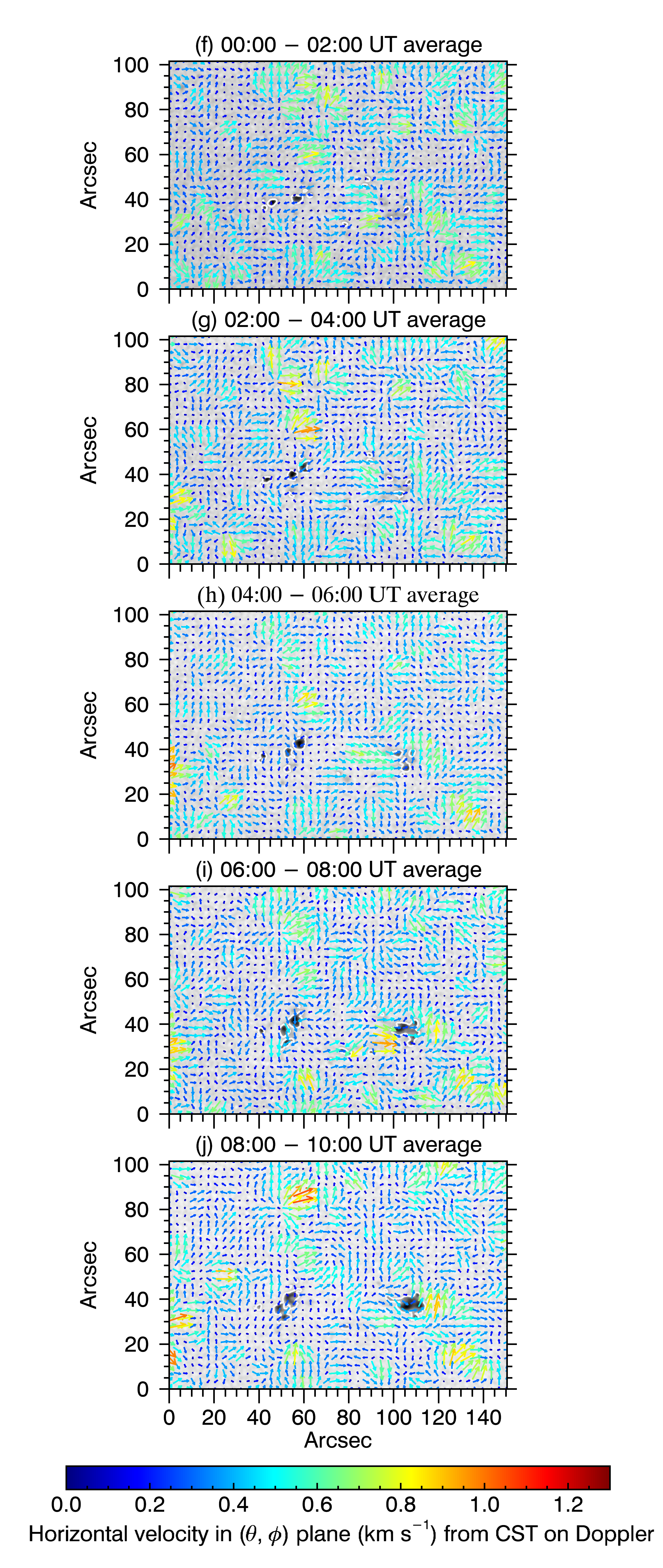}
	\caption{Horizontal velocity vector ($V_{\phi}$, $V_{\theta}$) 
	derived by applying CST to intensity (left column) and Doppler (right 
	column) maps observed by SDO/HMI on 29th of May 2023. Field of view 
	covers the plage region with emerging pores near to the 
	equator. The 
	top to bottom rows correspond to consecutive 120 min time 
	averages of the first 10 hours of observed data. Color and length of 
	the arrows indicate the horizontal velocity in a spectral scale 
	depicted by the color bar. The time averaged intensity image is 
	displayed in the background. The white contours mark the boundary 
	of the umbra of the pores. 
	}
\label{int_dop_hvel_29may2023}
\end{figure*}

Finally, we apply the CST algorithm to the first 10 hours of SDO/HMI 
observations on 29th of May 2023, which again corresponds to magnetically more 
active Sun. We particularly focus on studying the surface horizontal flows in a 
plage region with emerging pores near to the equator. For this purpose, 
we consider a field of view of $151 \times 102$ arcseconds covering the plage 
region. In this region, there exist two groups of pores (figure not 
illustrated), one to the left (east) and the other to the right (west). 
As the time progresses, we observe the continuous emergence of new pores 
between this existing group of pores embedded in the plage region, which is 
then followed by the growth of these pores into larger spots. We also notice  
that the group of pores to the left remains more or less at the same 
location for the entire 10 hours of observations, while those to the right 
moves towards the west. 

We have calculated the different correlation coefficients between 
horizontal velocities derived from the application of CST to intensity and 
Doppler maps around the plage region for the three different time 
windows, covering the 10-hour evolution of the pores. 
We find that over the entire course of evolution of the pores, the 
different correlation coefficients for the different flow quantities and time 
windows in a plage region are more or less similar to those 
obtained for the less active Sun case. As in the previously discussed cases, 
the different correlation coefficients increase with increasing time window. 
The MSE on the other hand shows larger variation with evolution of the 
pores, but remains 
below $0.055$, $0.035$, and $0.02$ for the 30, 60, and 120 min time window, 
respectively. Thus, we conclude 
that CST can be applied to Doppler maps around a plage region to derive 
meaningful horizontal flows when time averaging window is 60 min or larger. 

A comparison of horizontal velocities $(V_{\phi}, V_{\theta})$ 
determined from CST on intensity (left panels) and Doppler (right panels) 
maps around the plage region is shown in Figure~\ref{int_dop_hvel_29may2023} 
for 120 min time window successively for the 10 hours of observations. 
As in the case of sunspots (see Figure~\ref{int_dop_hvel_28may2023}), the 
horizontal velocities obtained from CST on Doppler are slightly larger in 
amplitude than those obtained from CST on intensity in the plage region. 
Horizontal flows around the pores have been studied by \citet{rketal23}. 
As reported by these authors, we also find inflows at the boundary 
(white contour) of only the left pores. In between the pores, we find  
horizontal flows from east to west around 75 to 80 arcseconds in the 
east-west direction and 25 to 35 arcseconds in the north-south direction. 
Further converging horizontal flows directed towards the west are seen on the 
left boundary of the west pores whose amplitude increases as the pores evolve 
and reaches a maximum for time average from 6 to 8 UT (see panels (d) and (i)) 
and then decreases (see panels (e) and (j)). Clearly, these relatively large 
amplitude converging horizontal flows 
are responsible for the movement of west pores toward the west with time. 
However, at the right edge of the west pores, we see horizontal flows pointing 
more towards north direction. All the above noted flows in and around the 
pores are recovered from both CST on intensity and Doppler.

\section{Conclusions}
\label{sec-conclu}
Coherent structure tracking (CST) is a tool to determine the solar 
surface dynamics, which is crucial to understand the solar magnetic 
fields. Since its inception by \citet{tretal99} and further 
developments by \citet{cstnew07,tretal12,tretal18}, CST has been 
traditionally applied to the intensity images of the Sun obtained 
in the continuum. In this paper, we apply, for the first time, the CST to
Dopplergrams. In particular, we consider Dopplergrams obtained 
from SDO/HMI and the vertical velocity ($V_z$) maps from a numerical 
simulation of granulation. We make a detailed comparison of the horizontal 
velocity fields determined by applying the CST to intensity and Dopplergrams 
and/ or vertical velocity maps. For this purpose, we compute the 
Pearson's linear and Spearman's and Kendall's rank-order 
correlation coefficients (both global and local) and also the 
MSE to measure the image similarity pixel by pixel. We also present 
histograms of velocity amplitude, scalar product 
and velocity differences between the horizontal velocities 
obtained from CST on intensity and that on Doppler. Furthermore, we 
compare the derivatives of the horizontal fields (both divergence and curl). 

We show that there is a high degree of similarity between the 
horizontal fields (in direction as well as amplitude) derived by the 
application of CST to intensity and to vertical velocity maps from a 
numerical simulation. In particular, we obtain a Pearson's linear correlation 
coefficient of 90\% or above for the horizontal velocities, and 80\% or 
above for the derivatives of the horizontal fields when 30 min time 
averaging is used (see Table~\ref{simu-table}). Spearman's rank-order 
correlation coefficient is more or less similar to the Pearson's linear 
correlation coefficient, while the Kendall's rank-order correlation is 
slightly smaller. However, all the three correlation coefficients 
increase with increasing time averaging window. Furthermore the MSE 
also indicated a high similarity (see Section~\ref{sec-comp-simu}), validating 
the application of CST to vertical velocity maps to derive horizontal 
flows.

As for the observations from SDO/HMI, we consider a relatively less 
active Sun data of 29th of December 2022 and magnetically more active Sun 
data of 28th and 29th of May 2023. In the case of relatively less active 
Sun, we obtain for the horizontal velocity fields of 30 minutes time 
resolution, a Pearson's linear GCC of 73\% and a near disk center LCC  
of 80\%. The derivative of the horizontal field, particularly the 
divergence also exhibits similar Pearson's linear correlation 
coefficient, while the curl being more noisy exhibits somewhat reduced 
correlation coefficient (see Tables~\ref{gcc-table} and 
\ref{lcc-table} and also Fig.~\ref{dopvsint-lcc}). When the time window 
for time averaging is increased these correlation coefficients also increase. 
Spearman's and Kendall's rank-order correlation coefficients exhibit 
a similar trend. However, they are somewhat smaller in value than the 
Pearson's linear correlation coefficient. MSE is also reasonably smaller in 
all the cases indicating a high similarity. The different correlation 
coefficients slightly decrease for the case of magnetically more active Sun 
with sunspots (see Table~\ref{lcc-sunspot-table}) or plage region 
with emerging pores, 
and also the MSE is relatively larger. Nevertheless we show that in the case 
of magnetically more active Sun, the meaningful horizontal flows can be 
obtained from CST on Doppler when time averaging window is 60 min or more, 
while a 30 min time window is sufficient for relatively less active Sun case.

We study the horizontal flows in the region covering the sunspots 
and a plage with emerging pores (see Figures~\ref{int_dop_hvel_28may2023} and 
\ref{int_dop_hvel_29may2023}). For this purpose, we plot the longitudinal 
($V_{\phi}$) and latitudinal ($V_{\theta}$) flows. We recover moat flows 
around the leading sunspot from CST on Doppler. Further, we study the evolution 
of horizontal flows in and around the pores, and show that the relatively 
large amplitude horizontal flows are responsible for the movement of right 
(west) pores towards the west. In general we find that the horizontal flows 
derived from CST on Doppler are slightly larger in amplitude than those 
obtained from CST on intensity. This may be due to the fact that HMI Doppler 
signals form at a slightly higher height in the photosphere than the HMI 
continuum intensity images (as the Doppler signals are derived using the 
Doppler shifts in the Fe\,{\sc i} 6173\ \AA\ absorption line). Thus, the 
CST on Doppler provides horizontal flows at heights slightly different 
than those obtained from CST on intensity.

With regard to the computational requirements, we note that CST on 
Dopplergrams completely avoids processing intensity images. 
This results in considerable savings in data storage, not 
only in the CST application but also when recording data from space- or 
ground-based instruments. Furthermore, the computing time 
is also somewhat reduced. 
More specifically, CST on Dopplergrams leads 
to about 9.4\% saving in CPU time when compared to the CST on intensity 
(including both the IDL and FORTRAN parts). The reason for reduction 
in time in CST-IDL part is that CST on Doppler only processes the 
Dopplergrams, while the CST on intensity processes both the continuum 
intensity images and the Dopplergrams (to derive the line-of-sight velocity). 
With regard to reduction in time in CST FORTRAN part, it is related to the 
fact that in intensity CST tends to detect more the smallest granules than in 
Doppler. The smaller granules seen in the intensity images do have a 
Doppler counterpart but with smaller amplitude. This is because, the smaller 
granules have a lower vertical velocity and hence their contrast in Doppler 
maps is also lower. Since the binarization or segmentation of the granules 
(both in intensity and Doppler maps) is based on local derivative and contrast 
\citep[see Sect.~2.1 of][]{cstnew07}, CST on Doppler tend to detect less 
number of smallest granules than the CST on intensity at the disk center. 
When we move to the limb due to further decrease in contrast, smaller granules 
tend to be lost in the CST segmented images, and hence they are not detected 
well by the CST. For example, for a 30 minutes time window CST on full-disk 
Doppler of a relatively less active Sun detects a total of about 20,000 
granules less than CST on intensity. With decreased detection of granules 
by CST on Doppler, the computing time for further operations such as granule 
identification and tracking is relatively reduced, leading to a saving in CPU 
time as mentioned above. However, despite this, in the present paper we show 
that the detected granules are sufficient to measure the horizontal flow 
fields. 

Clearly, our studies show 
that we can apply CST technique to Dopplergrams to derive the horizontal 
velocity fields on the surface of the Sun with the same level of confidence 
as CST on intensity images. Finally, we again emphasize that the 
main purpose of using CST on Doppler maps is to save space memory and time, 
even if it is just to save time. This becomes particularly critical 
when CST is applied to a longer time sequence of data, say over 
the solar cycle, to determine the meridional flows.

\begin{acknowledgements}
  We acknowledge the use of the high-performance computing facility 
 (\url{https://www.iiap.res.in/?q=facilities/computing/nova}) at the 
	Indian Institute of Astrophysics. We thank the SDO/HMI Science 
	Investigation Team for the free access to the data distributed through 
	JSOC. This work was supported by COFFIES, NASA Grant 80NSSC20K0602.
	M.S. would like to thank IRAP, OMP for providing the financial 
	support to visit IRAP for a period of 15 days in March 2024 when this 
	work was initiated. We thank the anonymous referee for 
	her/his insightful comments, which have improved the paper 
	considerably.
\end{acknowledgements}

\bibliographystyle{aa}

\begin{thebibliography}{32}
\expandafter\ifx\csname natexlab\endcsname\relax\def\natexlab#1{#1}\fi

\bibitem[{{Asensio Ramos} {et~al.}(2017){Asensio Ramos}, {Requerey}, \&
  {Vitas}}]{rametal17}
{Asensio Ramos}, A., {Requerey}, I.~S., \& {Vitas}, N. 2017, \aap, 604, A11

\bibitem[{{DeGrave} {et~al.}(2014){DeGrave}, {Jackiewicz}, \&
  {Rempel}}]{degetal14}
{DeGrave}, K., {Jackiewicz}, J., \& {Rempel}, M. 2014, \apj, 788, 127

\bibitem[{{Duvall} {et~al.}(1993){Duvall}, {Jefferies}, {Harvey}, \&
  {Pomerantz}}]{djh93}
{Duvall}, Jr., T.~L., {Jefferies}, S.~M., {Harvey}, J.~W., \& {Pomerantz},
  M.~A. 1993, \nat, 362, 430

\bibitem[{{Fisher} \& {Welsch}(2008)}]{fw08}
{Fisher}, G.~H. \& {Welsch}, B.~T. 2008, in Astronomical Society of the Pacific
  Conference Series, Vol. 383, Subsurface and Atmospheric Influences on Solar
  Activity, ed. R.~{Howe}, R.~W. {Komm}, K.~S. {Balasubramaniam}, \& G.~J.~D.
  {Petrie}, 373

\bibitem[{{Kamlah} {et~al.}(2023){Kamlah}, {Verma}, {Denker}, \&
  {Wang}}]{rketal23}
{Kamlah}, R., {Verma}, M., {Denker}, C., \& {Wang}, H. 2023, \aap, 675, A182

\bibitem[{{Li} {et~al.}(2023){Li}, {Xu}, {Verma}, {Denker}, {Zhao}, \&
  {Wang}}]{lietal23}
{Li}, Q., {Xu}, Y., {Verma}, M., {et~al.} 2023, \solphys, 298, 62

\bibitem[{{L{\"o}hner-B{\"o}ttcher} \& {Schlichenmaier}(2013)}]{LBS13}
{L{\"o}hner-B{\"o}ttcher}, J. \& {Schlichenmaier}, R. 2013, \aap, 551, A105

\bibitem[{{L{\"o}ptien} {et~al.}(2017){L{\"o}ptien}, {Birch}, {Duvall},
  {Gizon}, {Proxauf}, \& {Schou}}]{loptetal17}
{L{\"o}ptien}, B., {Birch}, A.~C., {Duvall}, T.~L., {et~al.} 2017, \aap, 606,
  A28

\bibitem[{{Louis} {et~al.}(2015){Louis}, {Ravindra}, {Georgoulis}, \&
  {K{\"u}ker}}]{louisetal15}
{Louis}, R.~E., {Ravindra}, B., {Georgoulis}, M.~K., \& {K{\"u}ker}, M. 2015,
  \solphys, 290, 1135

\bibitem[{{Mahajan} {et~al.}(2024){Mahajan}, {Upton}, {Antia}, {Basu},
  {DeRosa}, {Hess Webber}, {Hoeksema}, {Jain}, {Komm}, {Larson}, {Nagovitsyn},
  {Pevtsov}, {Roudier}, {Tripathy}, {Ulrich}, \& {Zhao}}]{mahaetal24}
{Mahajan}, S.~S., {Upton}, L.~A., {Antia}, H.~M., {et~al.} 2024, \solphys, 299,
  38

\bibitem[{{November}(1986)}]{nov86}
{November}, L.~J. 1986, \ao, 25, 392

\bibitem[{{Pesnell} {et~al.}(2012){Pesnell}, {Thompson}, \&
  {Chamberlin}}]{sdo12}
{Pesnell}, W.~D., {Thompson}, B.~J., \& {Chamberlin}, P.~C. 2012, \solphys,
  275, 3

\bibitem[{{Rieutord} {et~al.}(2001){Rieutord}, {Roudier}, {Ludwig}, {Nordlund},
  \& {Stein}}]{rieuetal01}
{Rieutord}, M., {Roudier}, T., {Ludwig}, H.-G., {Nordlund}, {\r{A}}., \&
  {Stein}, R. 2001, \aap, 377, L14

\bibitem[{{Rieutord} {et~al.}(2007){Rieutord}, {Roudier}, {Roques}, \&
  {Ducottet}}]{cstnew07}
{Rieutord}, M., {Roudier}, T., {Roques}, S., \& {Ducottet}, C. 2007, \aap, 471,
  687

\bibitem[{{Rincon} {et~al.}(2017){Rincon}, {Roudier}, {Schekochihin}, \&
  {Rieutord}}]{rinetal17}
{Rincon}, F., {Roudier}, T., {Schekochihin}, A.~A., \& {Rieutord}, M. 2017,
  \aap, 599, A69

\bibitem[{{Roudier} {et~al.}(2020){Roudier}, {Malherbe}, {Gelly}, {Douet},
  {Frank}, \& {Dalmasse}}]{tretal20}
{Roudier}, T., {Malherbe}, J.~M., {Gelly}, B., {et~al.} 2020, \aap, 641, A50

\bibitem[{{Roudier} {et~al.}(2019){Roudier}, {Malherbe}, {Stein}, \&
  {Frank}}]{tretal19}
{Roudier}, T., {Malherbe}, J.~M., {Stein}, R.~F., \& {Frank}, Z. 2019, \aap,
  622, A112

\bibitem[{{Roudier} {et~al.}(2012){Roudier}, {Rieutord}, {Malherbe}, {Renon},
  {Berger}, {Frank}, {Prat}, {Gizon}, \& {{\v{S}}vanda}}]{tretal12}
{Roudier}, T., {Rieutord}, M., {Malherbe}, J.~M., {et~al.} 2012, \aap, 540, A88

\bibitem[{{Roudier} {et~al.}(1999){Roudier}, {Rieutord}, {Malherbe}, \&
  {Vigneau}}]{tretal99}
{Roudier}, T., {Rieutord}, M., {Malherbe}, J.~M., \& {Vigneau}, J. 1999, \aap,
  349, 301

\bibitem[{{Roudier} {et~al.}(2013){Roudier}, {Rieutord}, {Prat}, {Malherbe},
  {Renon}, {Frank}, {{\v{S}}vanda}, {Berger}, {Burston}, \& {Gizon}}]{tretal13}
{Roudier}, T., {Rieutord}, M., {Prat}, V., {et~al.} 2013, \aap, 552, A113

\bibitem[{{Roudier} {et~al.}(2018){Roudier}, {{\v{S}}vanda}, {Ballot},
  {Malherbe}, \& {Rieutord}}]{tretal18}
{Roudier}, T., {{\v{S}}vanda}, M., {Ballot}, J., {Malherbe}, J.~M., \&
  {Rieutord}, M. 2018, \aap, 611, A92

\bibitem[{{Scherrer} {et~al.}(2012){Scherrer}, {Schou}, {Bush}, {Kosovichev},
  {Bogart}, {Hoeksema}, {Liu}, {Duvall}, {Zhao}, {Title}, {Schrijver},
  {Tarbell}, \& {Tomczyk}}]{hmi112}
{Scherrer}, P.~H., {Schou}, J., {Bush}, R.~I., {et~al.} 2012, \solphys, 275,
  207

\bibitem[{{Schou} {et~al.}(2012){Schou}, {Scherrer}, {Bush}, {Wachter},
  {Couvidat}, {Rabello-Soares}, {Bogart}, {Hoeksema}, {Liu}, {Duvall}, {Akin},
  {Allard}, {Miles}, {Rairden}, {Shine}, {Tarbell}, {Title}, {Wolfson},
  {Elmore}, {Norton}, \& {Tomczyk}}]{hmi212}
{Schou}, J., {Scherrer}, P.~H., {Bush}, R.~I., {et~al.} 2012, \solphys, 275,
  229

\bibitem[{{Stein}(2012)}]{s12}
{Stein}, R.~F. 2012, Living Reviews in Solar Physics, 9, 4

\bibitem[{{Stein} \& {Nordlund}(1998)}]{sn98}
{Stein}, R.~F. \& {Nordlund}, {\r{A}}. 1998, \apj, 499, 914

\bibitem[{{Stein} {et~al.}(2009){Stein}, {Nordlund}, {Georgoviani}, {Benson},
  \& {Schaffenberger}}]{setal09}
{Stein}, R.~F., {Nordlund}, {\r{A}}., {Georgoviani}, D., {Benson}, D., \&
  {Schaffenberger}, W. 2009, in Astronomical Society of the Pacific Conference
  Series, Vol. 416, Solar-Stellar Dynamos as Revealed by Helio- and
  Asteroseismology: GONG 2008/SOHO 21, ed. M.~{Dikpati}, T.~{Arentoft},
  I.~{Gonz{\'a}lez Hern{\'a}ndez}, C.~{Lindsey}, \& F.~{Hill}, 421

\bibitem[{{Strous}(1994)}]{sl94}
{Strous}, L.~H. 1994, PhD thesis, University of Utrecht, Netherlands

\bibitem[{{Strous}(1995)}]{sl95}
{Strous}, L.~H. 1995, in ESA Special Publication, Vol. 376, Helioseismology,
  ed. J.~T. {Hoeksema}, V.~{Domingo}, B.~{Fleck}, \& B.~{Battrick}, 213

\bibitem[{{Tremblay} {et~al.}(2018){Tremblay}, {Roudier}, {Rieutord}, \&
  {Vincent}}]{tremetal18}
{Tremblay}, B., {Roudier}, T., {Rieutord}, M., \& {Vincent}, A. 2018, \solphys,
  293, 57

\bibitem[{{Upton} {et~al.}(2024){Upton}, {Mahajan}, {Antia}, {Basu}, {Biji},
  {Hathaway}, {Hess Webber}, {Hoeksema}, {Jain}, {Komm}, {Jha}, {Lamb},
  {Pevtsov}, {Roudier}, {Tripathy}, {Ulrich}, \& {Zhao}}]{upetal24}
{Upton}, L., {Mahajan}, S., {Antia}, H.~M., {et~al.} 2024, in AGU Fall Meeting
  Abstracts, Vol. 2024, AGU Fall Meeting Abstracts, SH13A--2904

\bibitem[{{Verma} {et~al.}(2018){Verma}, {Kummerow}, \& {Denker}}]{vermaetal18}
{Verma}, M., {Kummerow}, P., \& {Denker}, C. 2018, Astronomische Nachrichten,
  339, 268

\bibitem[{{Verma} {et~al.}(2013){Verma}, {Steffen}, \& {Denker}}]{vermaetal13}
{Verma}, M., {Steffen}, M., \& {Denker}, C. 2013, \aap, 555, A136

\end{thebibliography}

\onecolumn
\begin{appendix}
\section{LCC across the full-disk of a relatively less active Sun}
\label{lccfig-app}

\begin{figure}[ht]
\centering 
\includegraphics[scale=0.32]{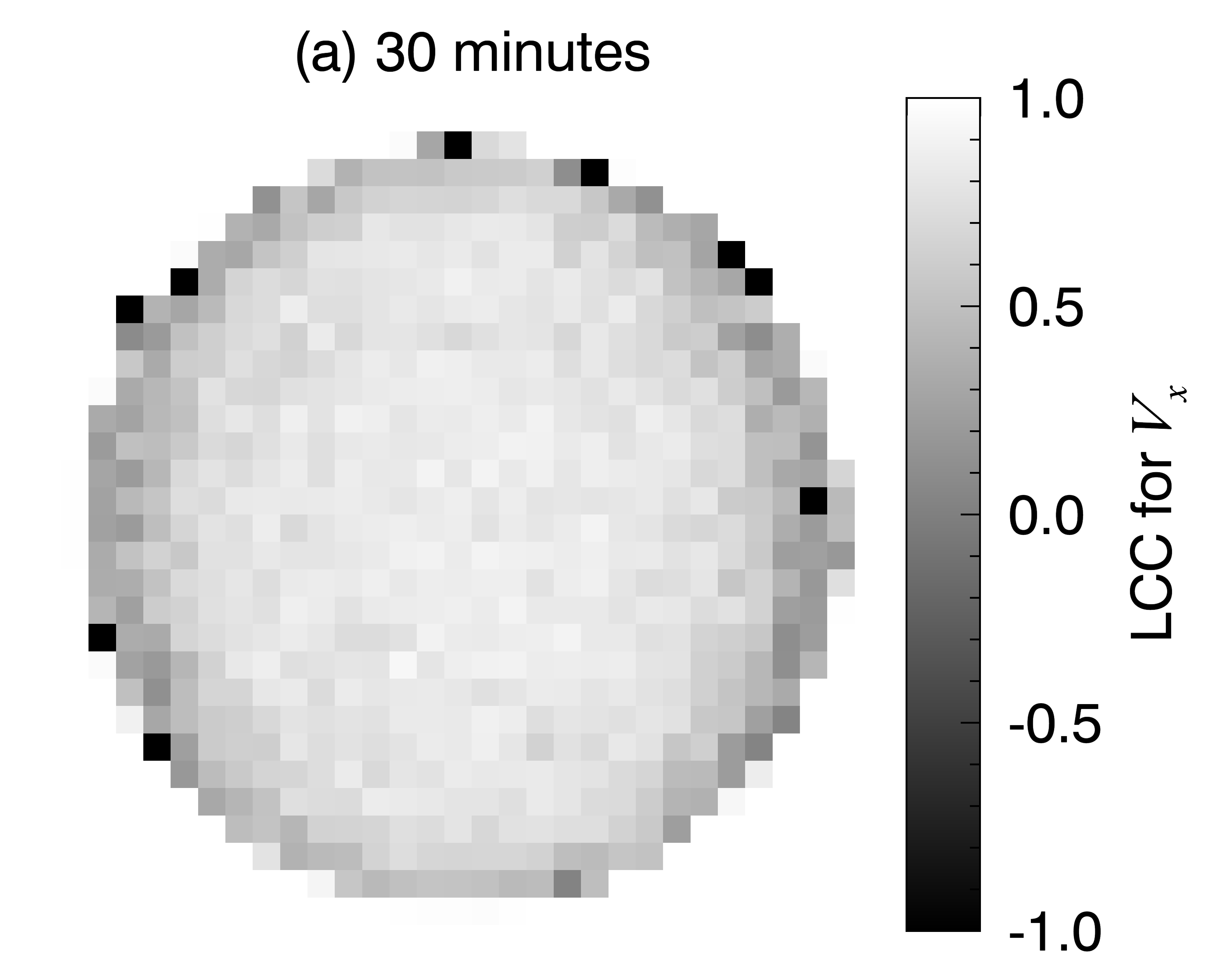}\ \ \ \ 
\includegraphics[scale=0.32]{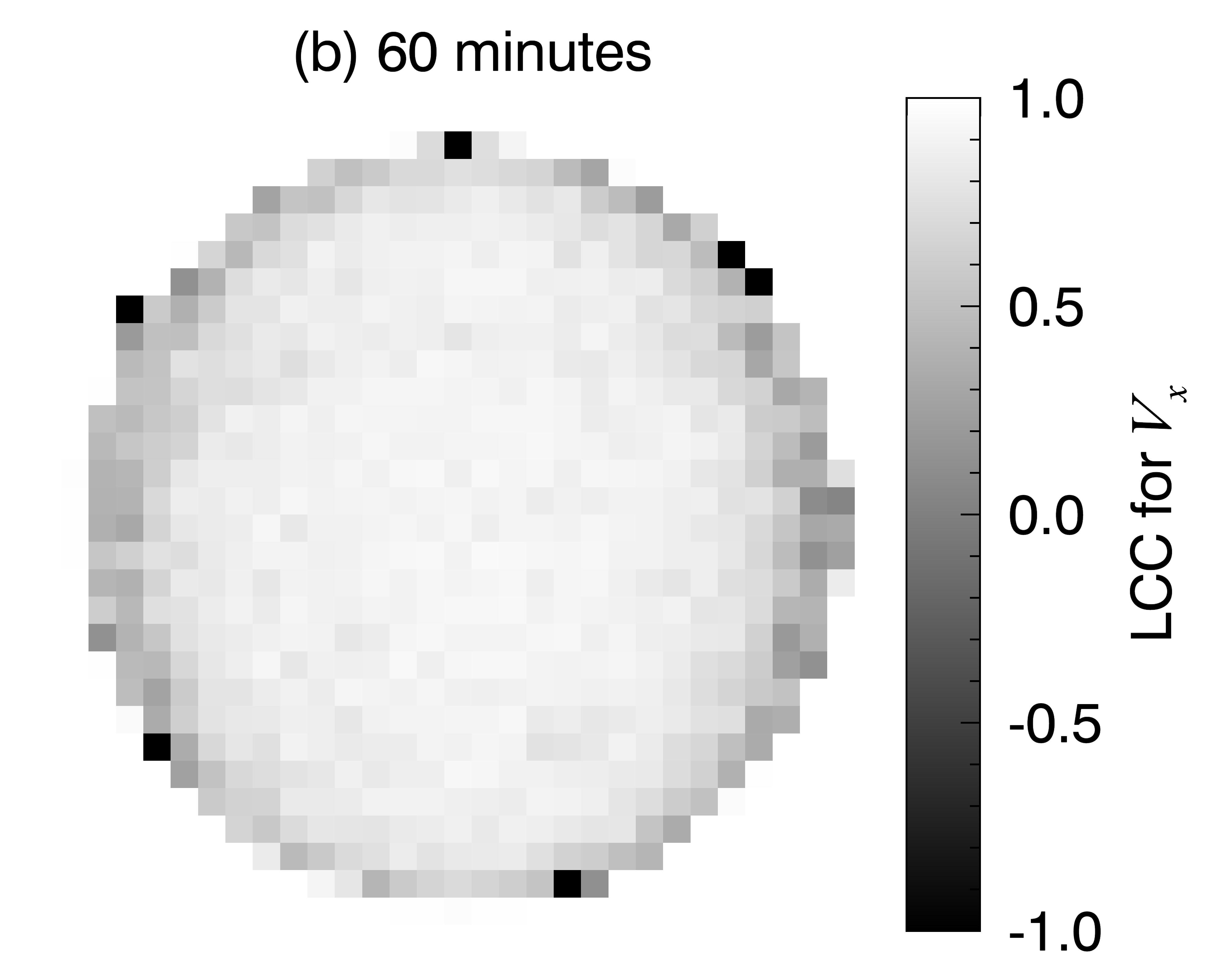}\ \ \ \ 
\includegraphics[scale=0.32]{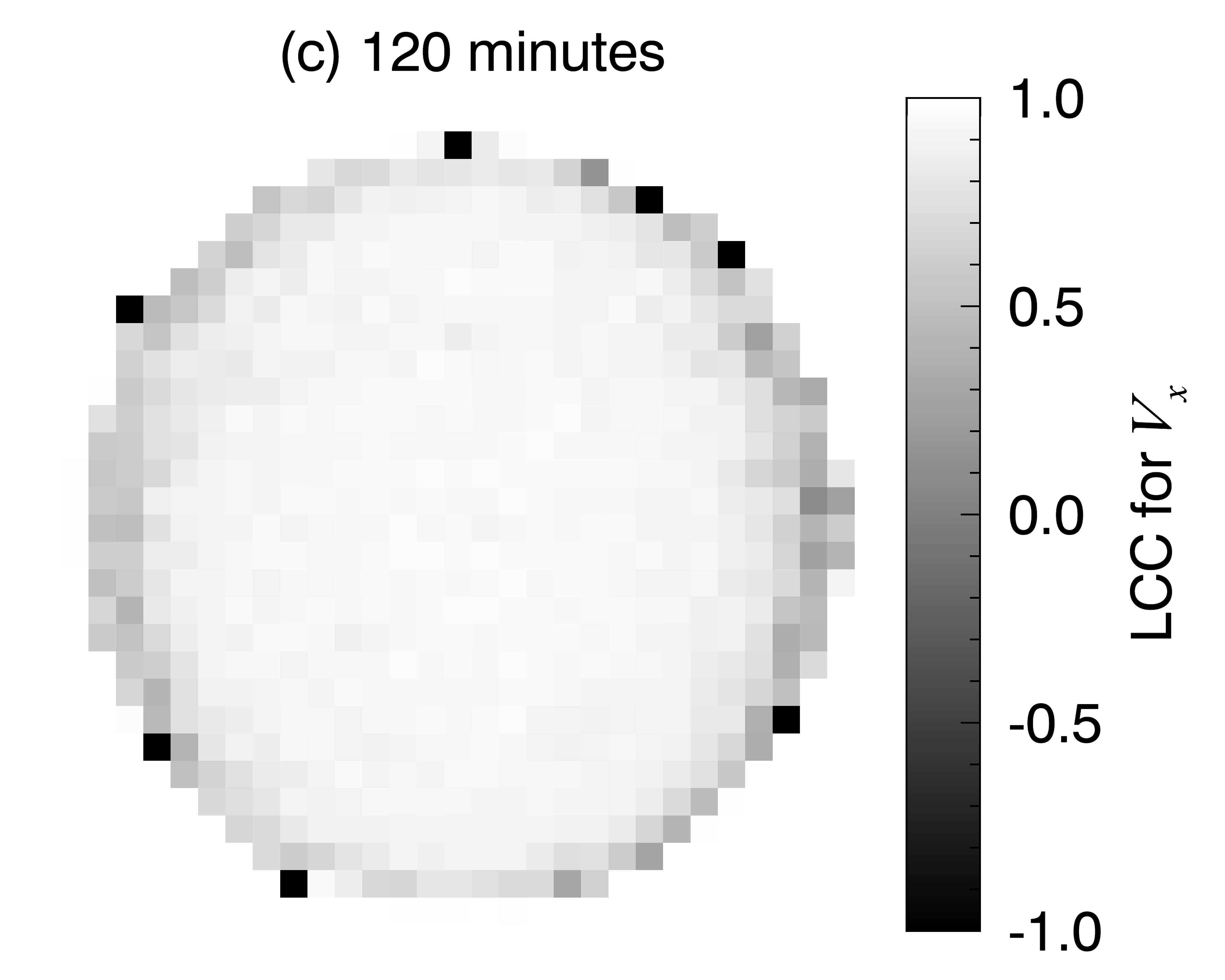}\\ 
\includegraphics[scale=0.32]{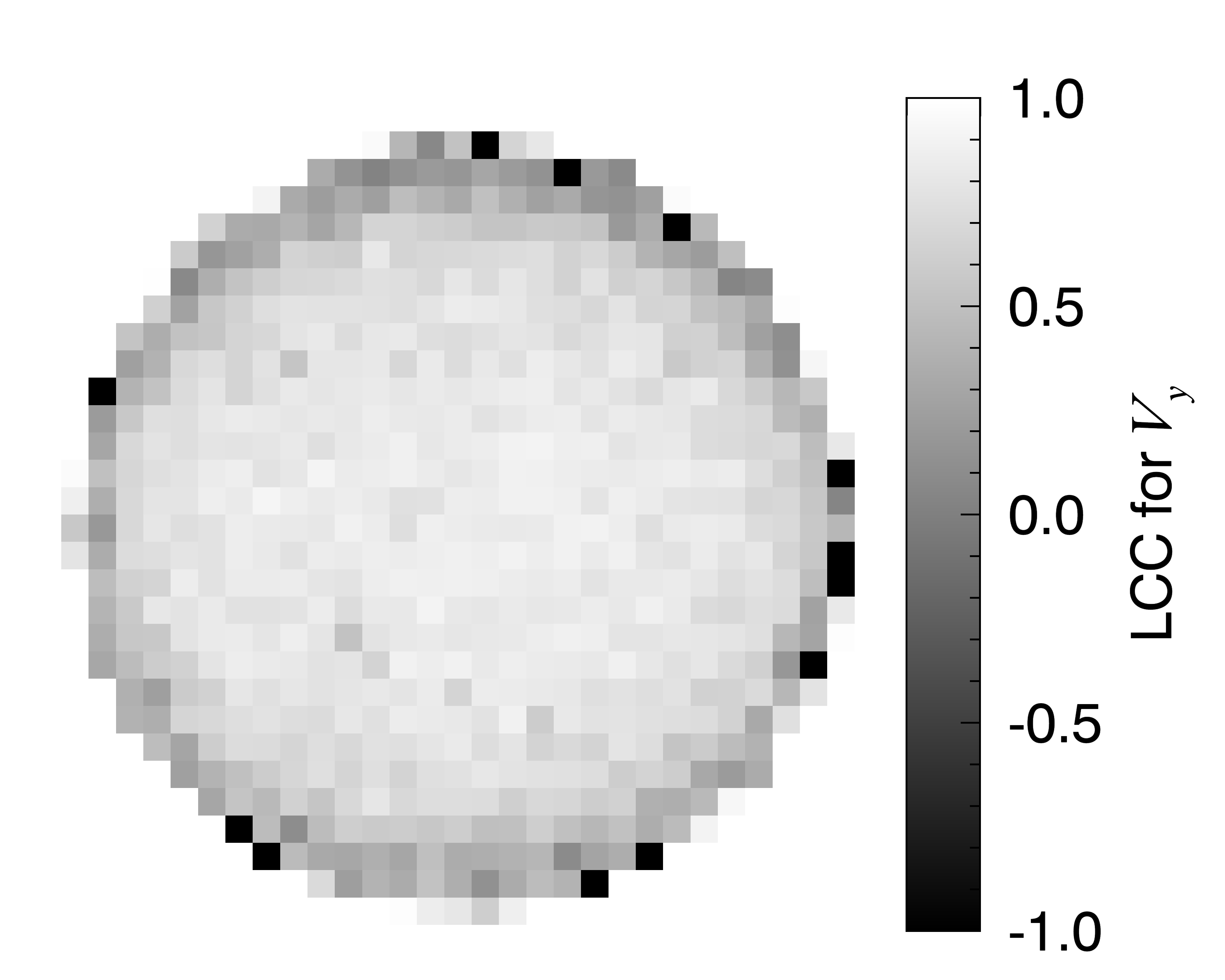}\ \ \ \ 
\includegraphics[scale=0.32]{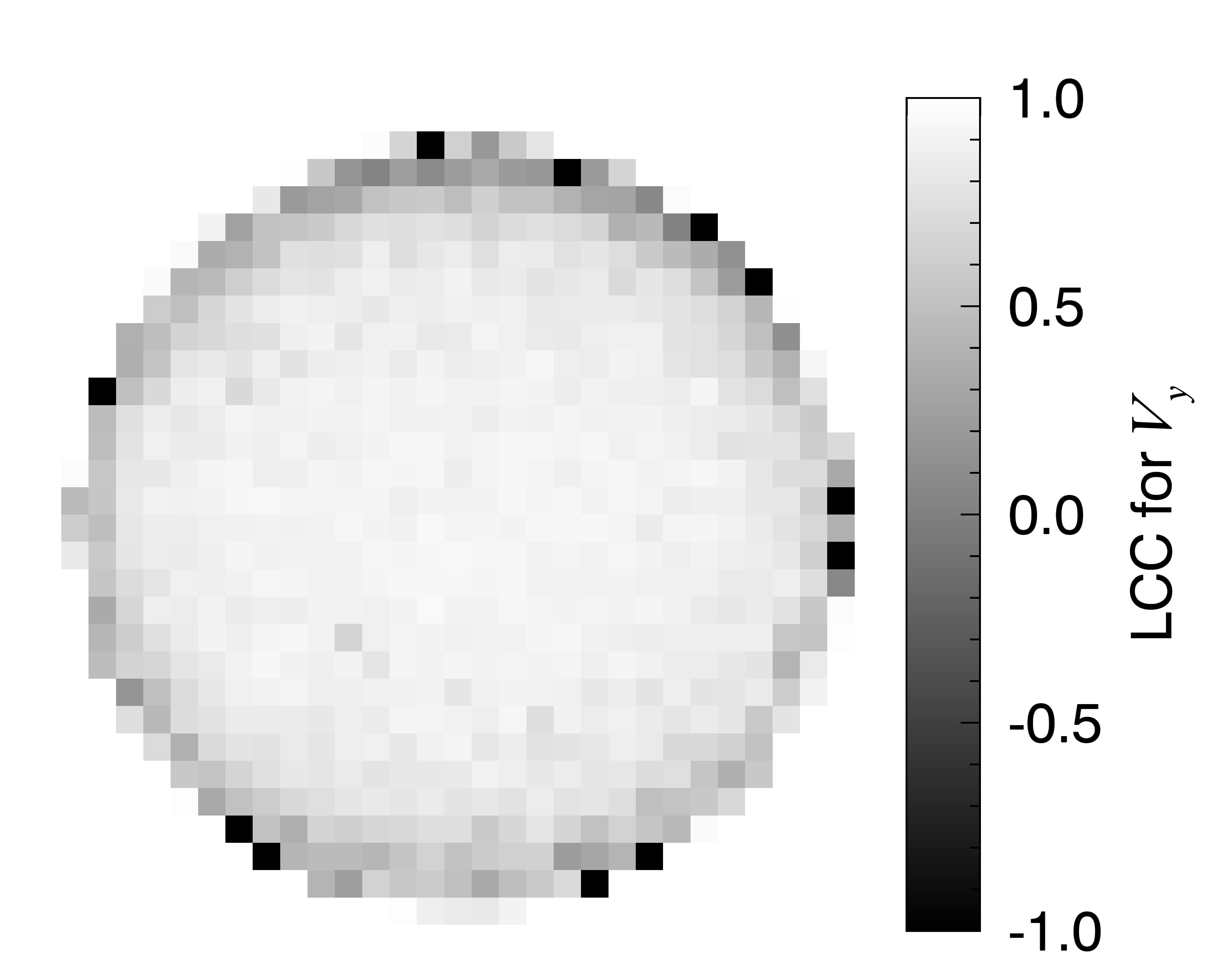}\ \ \ \ 
\includegraphics[scale=0.32]{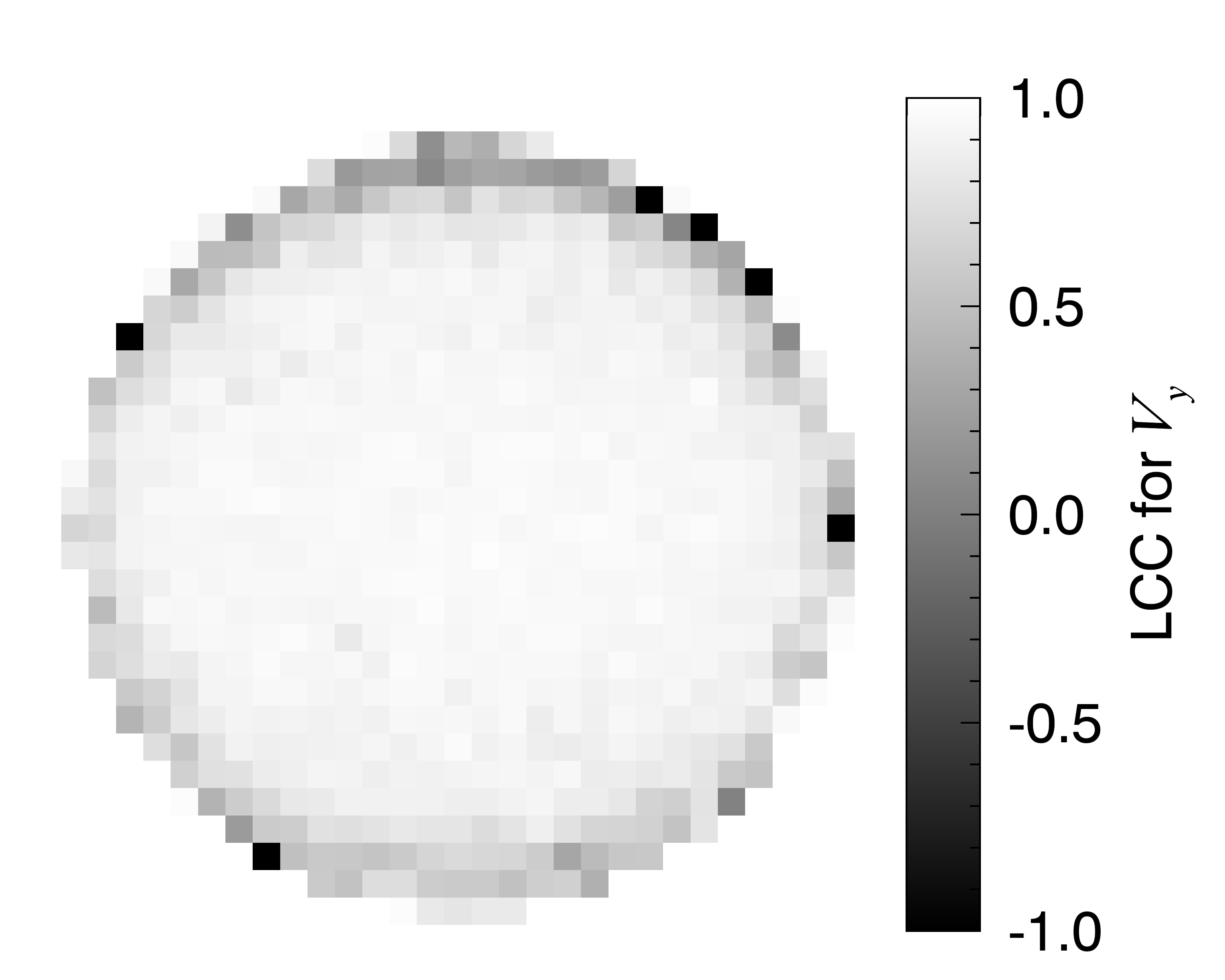}\\ 
\includegraphics[scale=0.32]{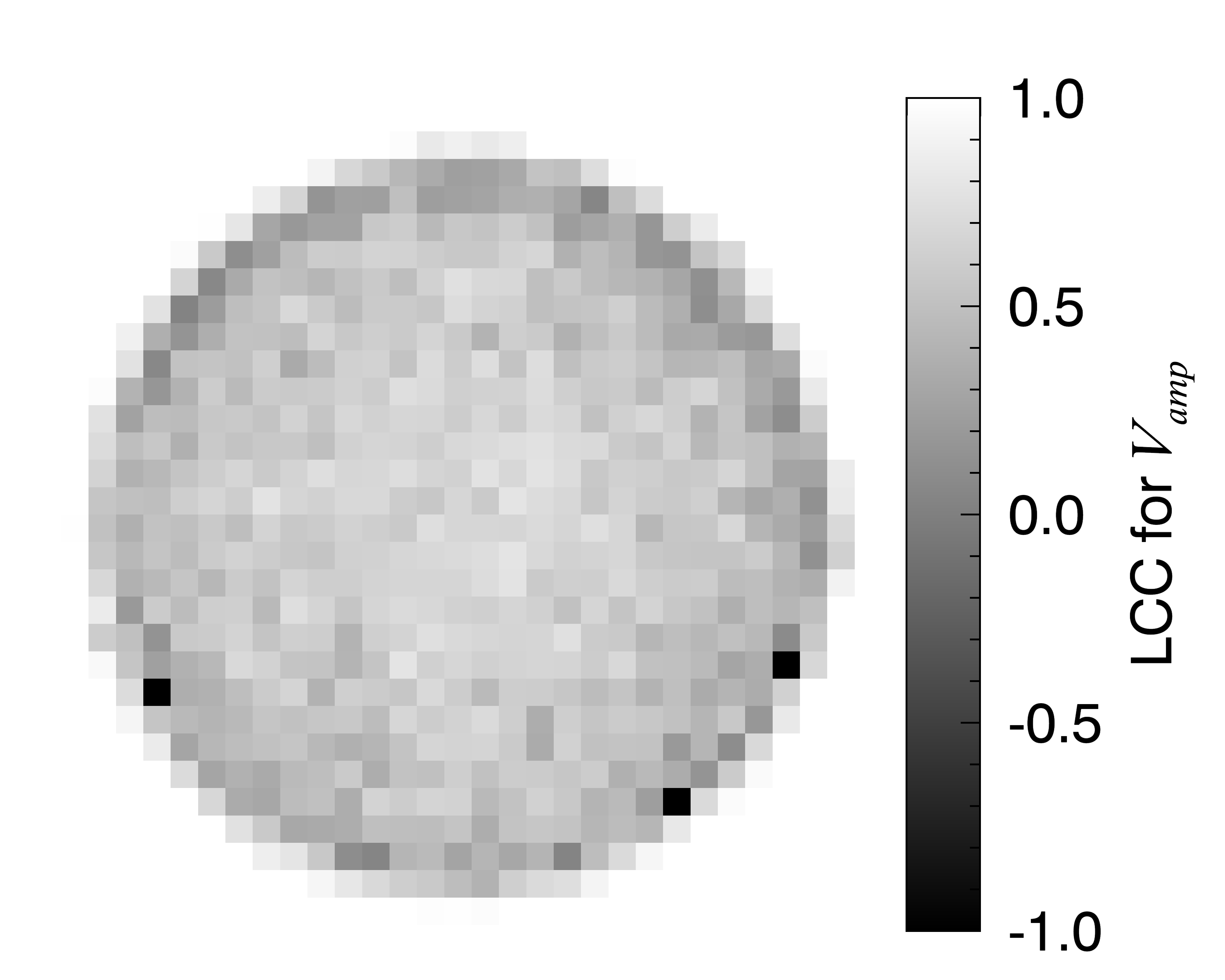}\ \ \ \ 
\includegraphics[scale=0.32]{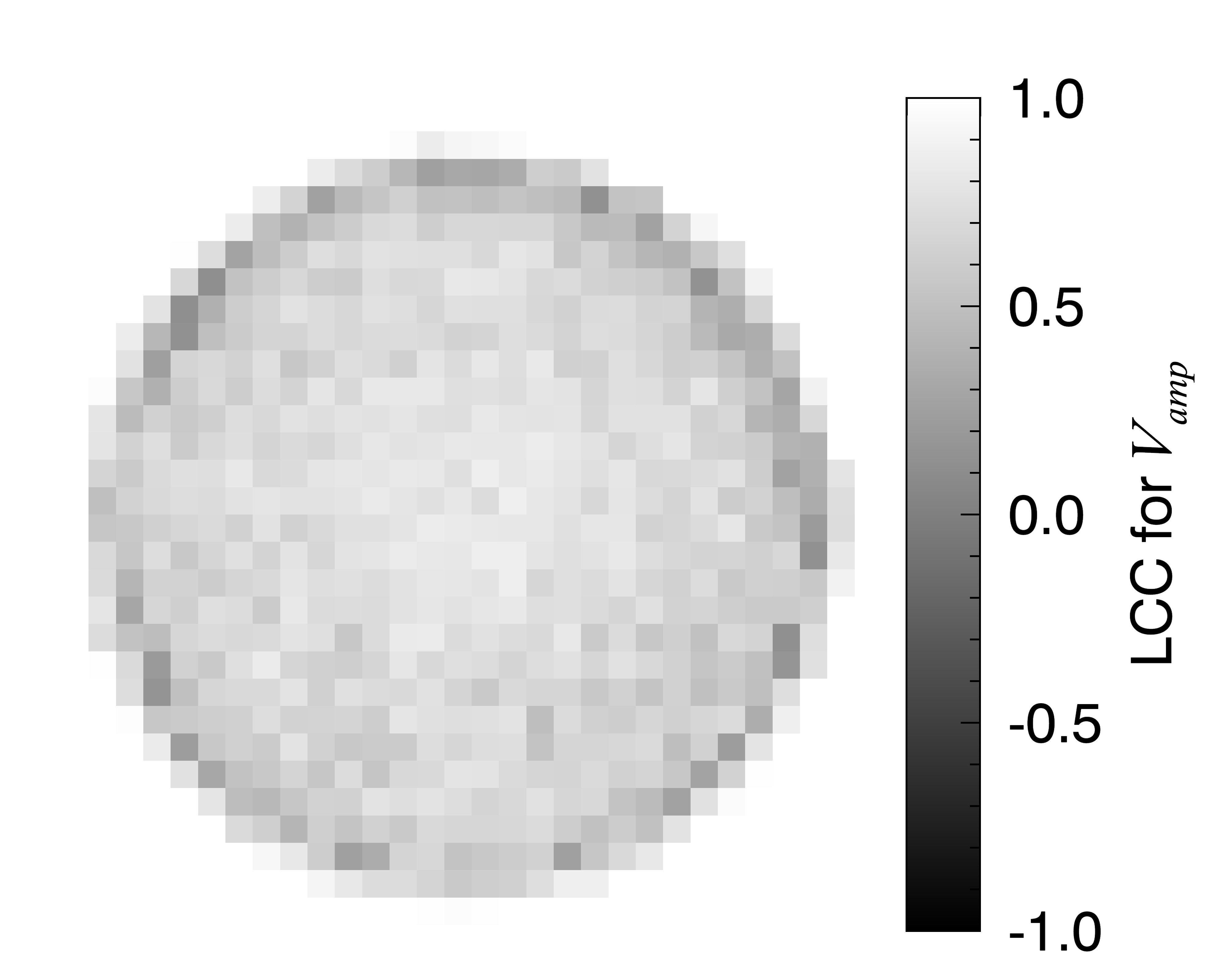}\ \ \ \ 
\includegraphics[scale=0.32]{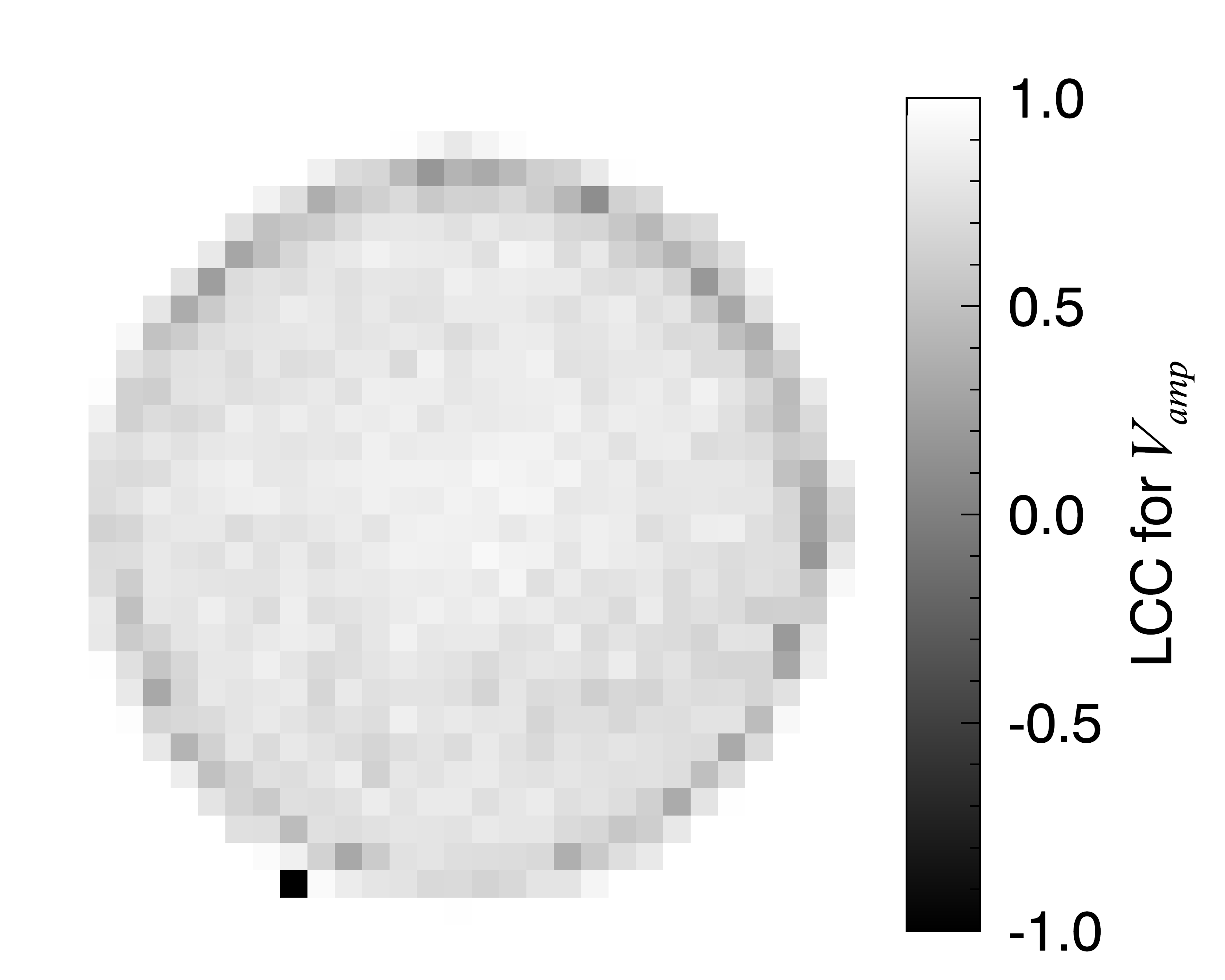}\\ 
\includegraphics[scale=0.32]{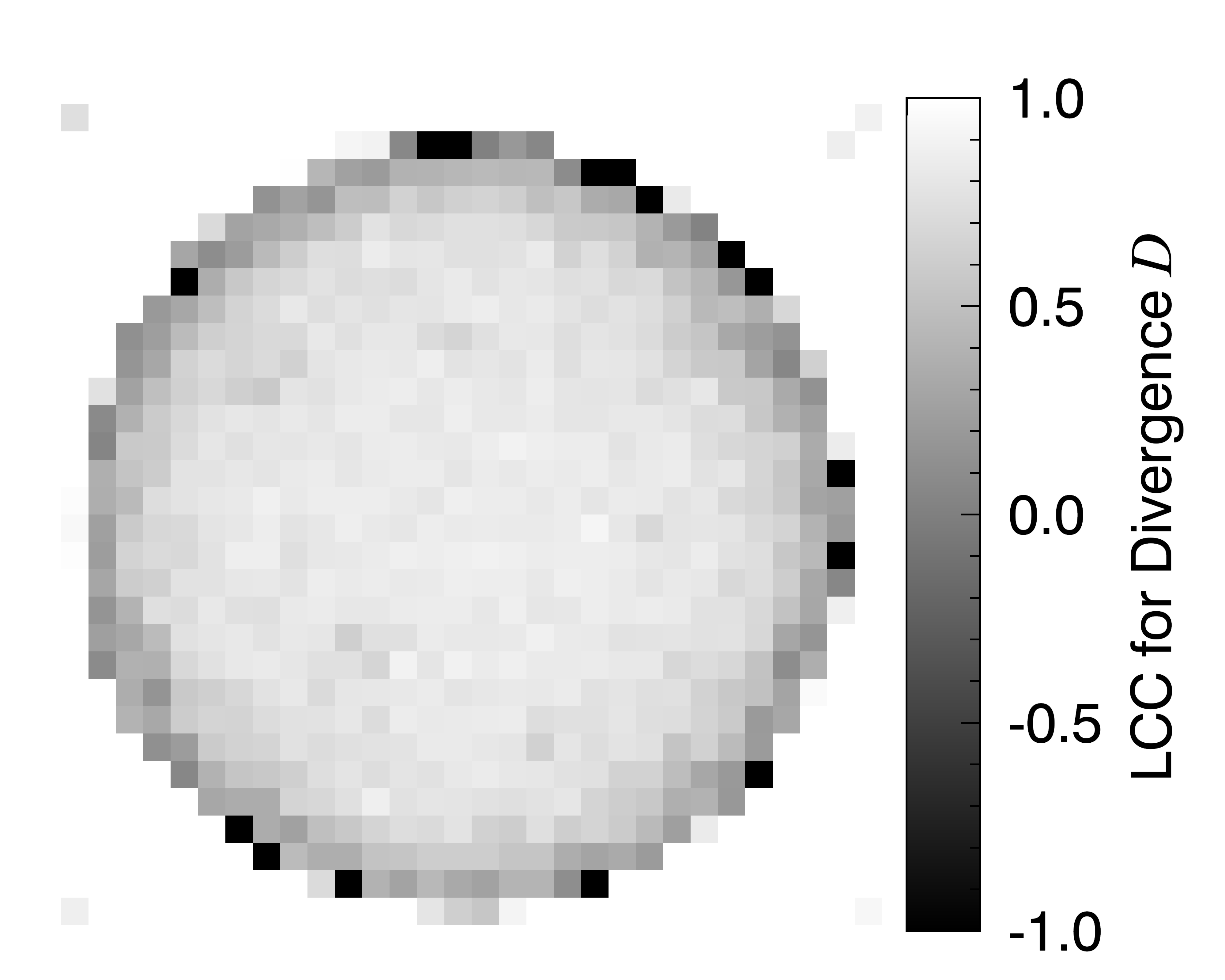}\ \ \ \ 
\includegraphics[scale=0.32]{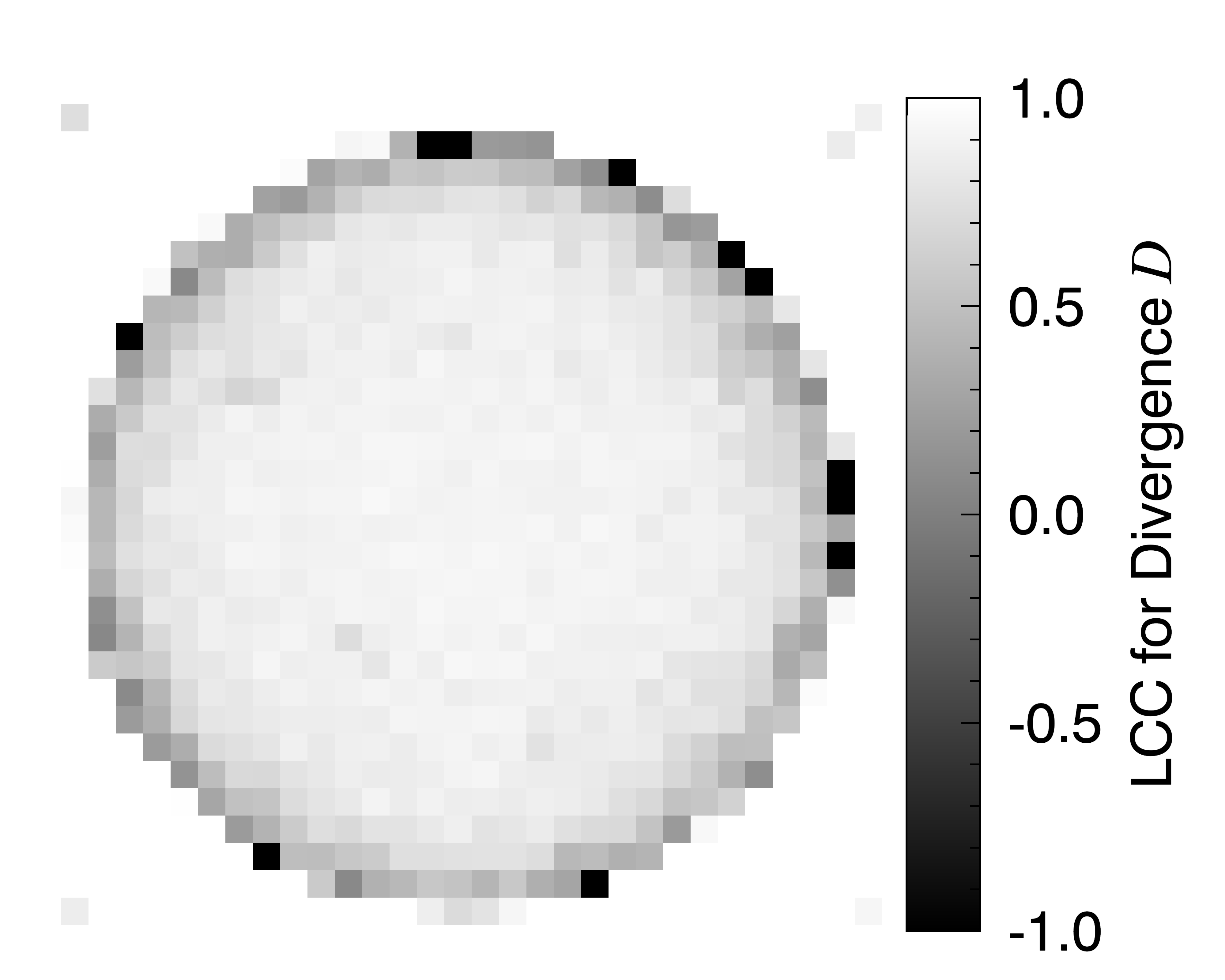}\ \ \ \ 
\includegraphics[scale=0.32]{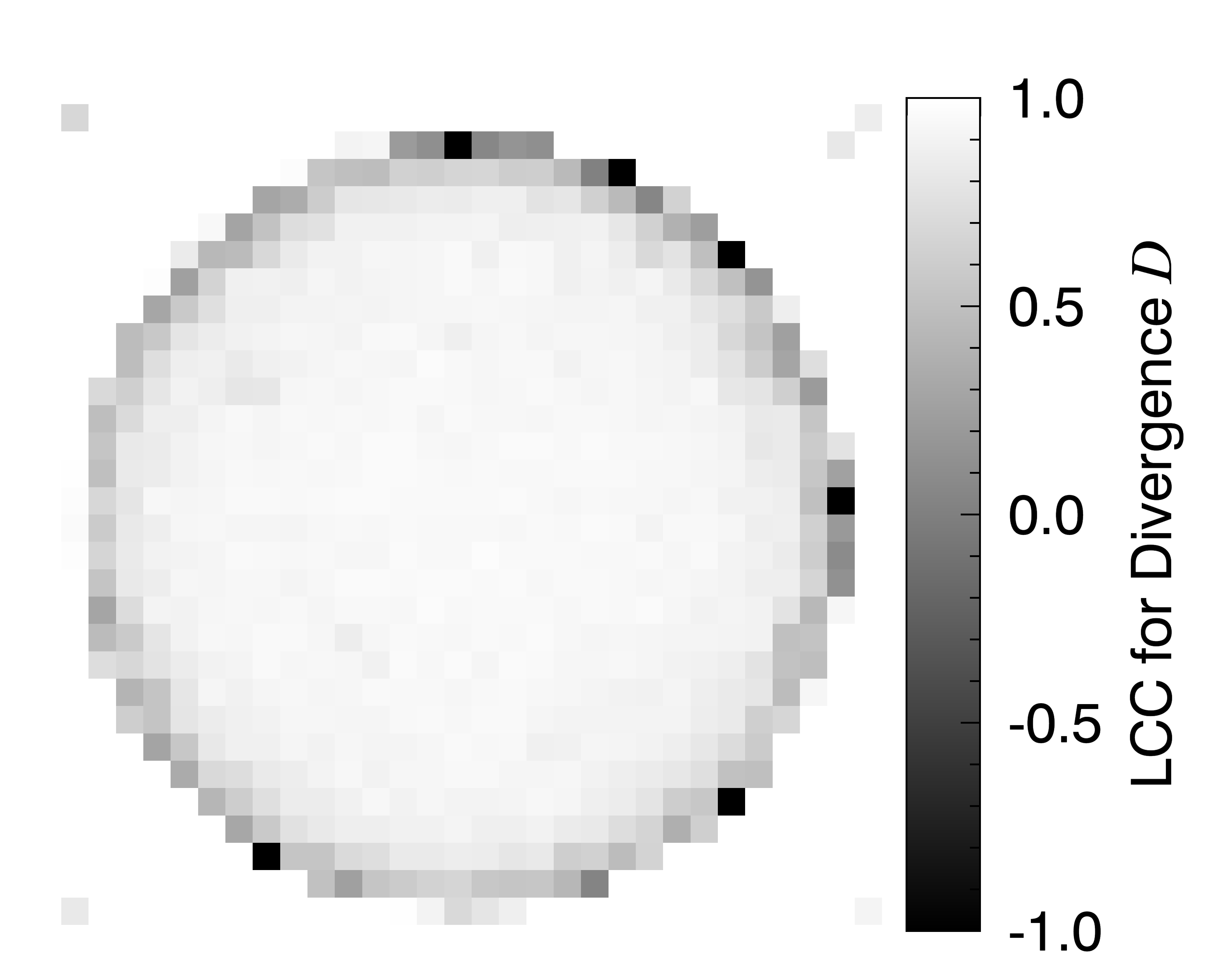}\\ 
\includegraphics[scale=0.32]{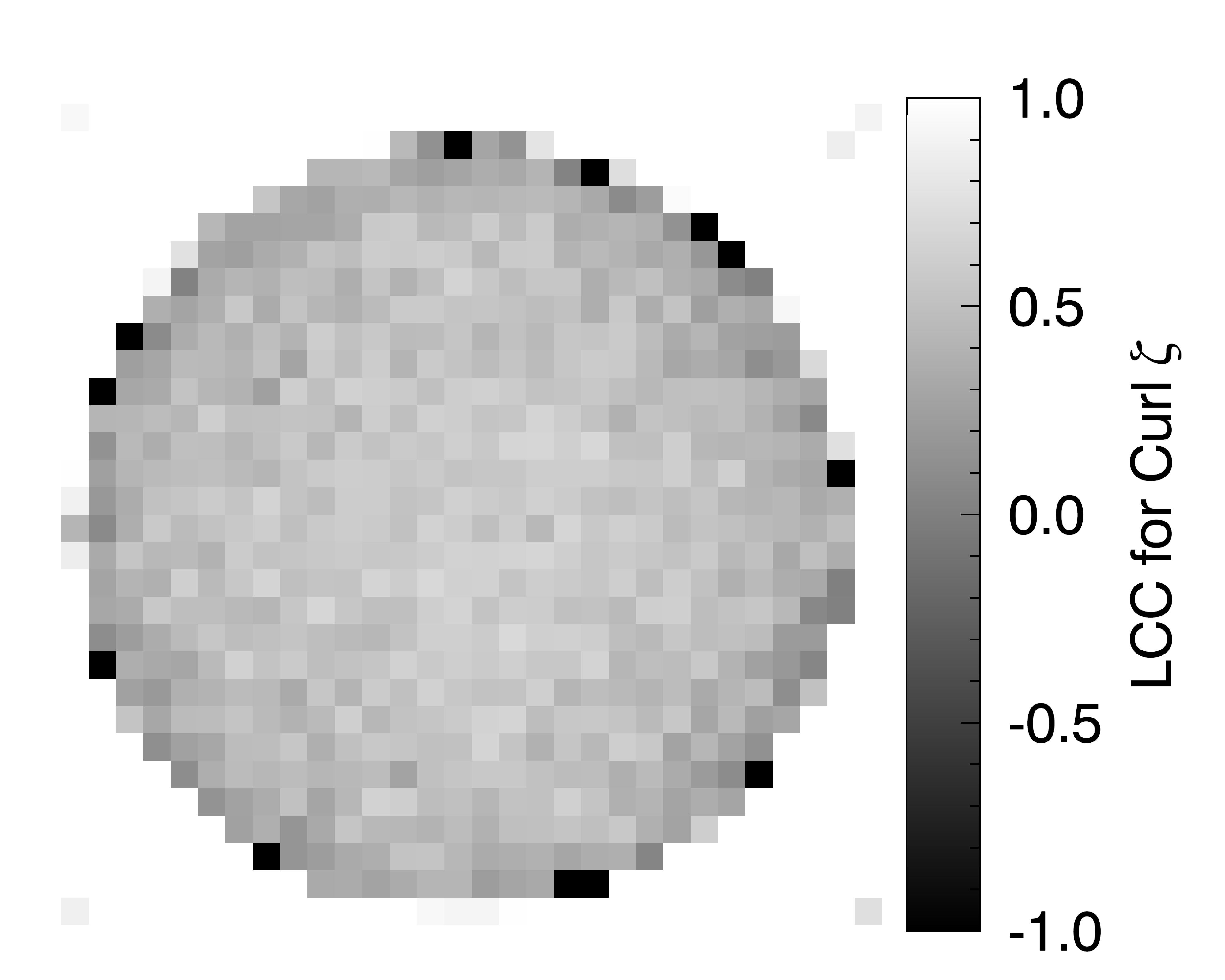}\ \ \ \ 
\includegraphics[scale=0.32]{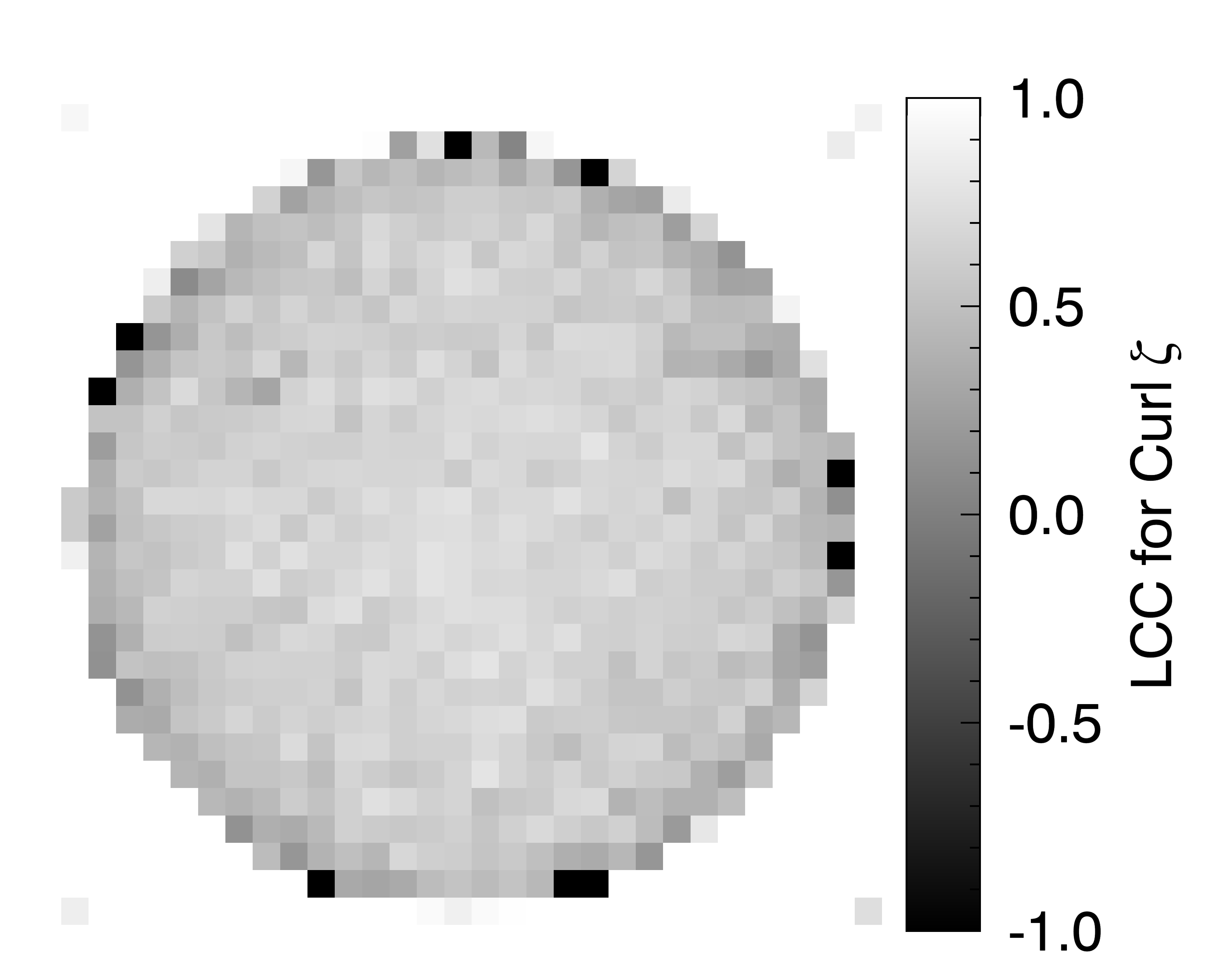}\ \ \ \ 
\includegraphics[scale=0.32]{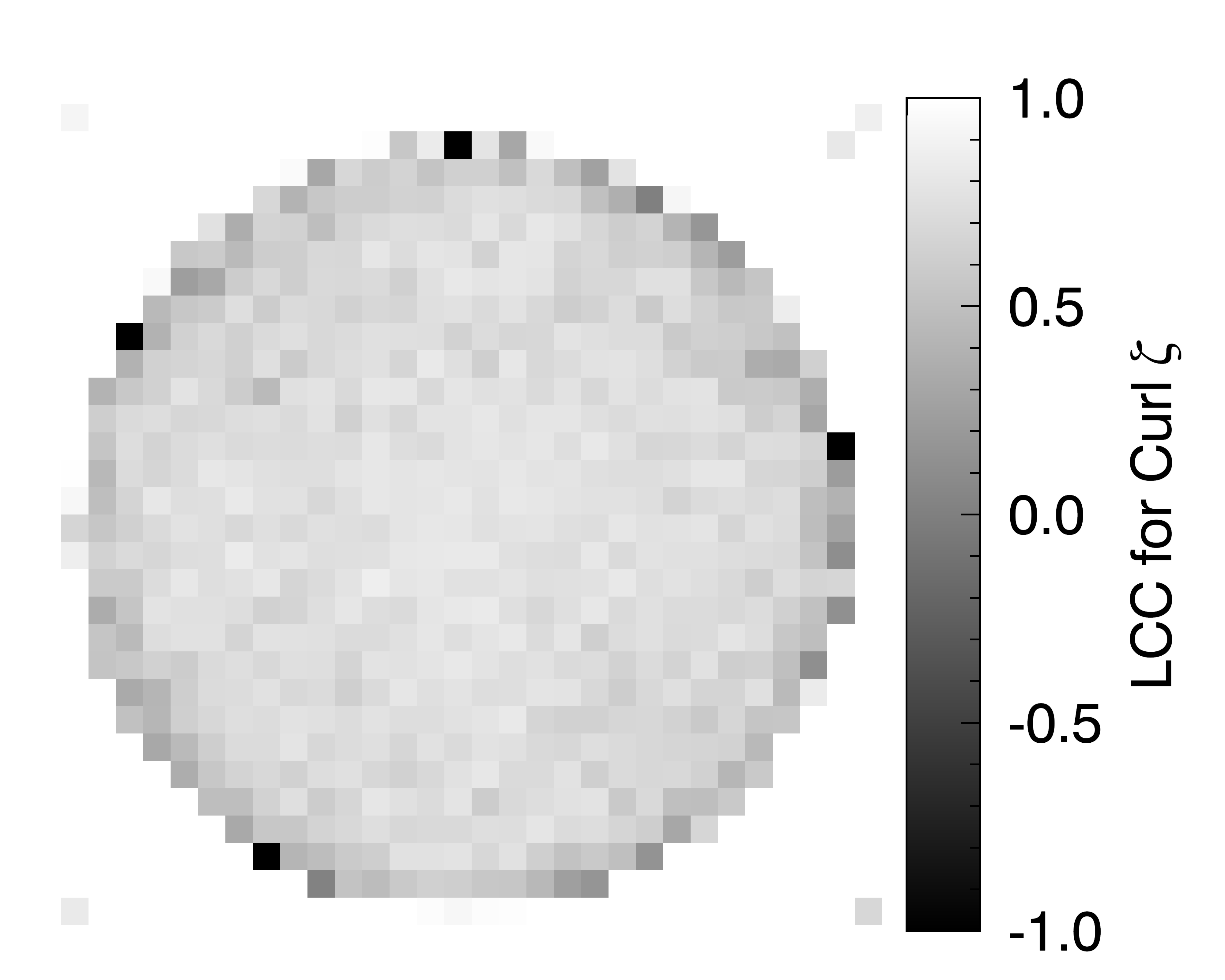}
	\caption{Pearson's linear LCC for different flow 
	quantities across the entire disk of the Sun and with the 
	30 (panels (a)), 60 (panels (b)), and 120 (panels (c)) minutes 
	time averaging. Relatively less active Sun observations from 
	SDO/HMI on 29th of December 2022 have been considered. Clearly for 
	all the flow quantities the LCC increases with the temporal window for 
	time averaging. }
  \label{dopvsint-lcc}
\end{figure}
\end{appendix}

\end{document}